\documentclass[12pt,a4paper]{article}
\usepackage{amssymb}
\usepackage{amsmath}
\usepackage{amsfonts}
\usepackage[margin=1.1in]{geometry}
\begin{document}

\title{Asymptotic properties of a Nadaraya-Watson type\\
estimator for regression functions of infinite order\thanks{%
Part of this paper was written while the first author visited Institut de
Math\'{e}matiques, Toulouse in November 2013. We thank Philippe Vieu for his
kind hospitality and many helpful advice. We also thank John Aston, Jeroen
Dalderop, Paul Doukhan, Jiti Gao, Mikhail Lifshits, Alexei Onatski, and
Richard Nickl for their helpful comments. Special thanks to Alexey Rudenko
for providing an original Russian photocopy of Sytaya (1974), and Hyungjin
Lee for translating the paper. Financial support from the European Research
Council (ERC-2008AdG-NAMSEF) is gratefully acknowledged.} }
\author{Seok Young Hong\thanks{%
Statistical Laboratory, Department of Pure Mathematics and Mathematical
Statistics, University of Cambridge, United Kingdom, email: syh30@cam.ac.uk}%
\quad~ \quad Oliver Linton\thanks{%
Faculty of Economics, University of Cambridge, United Kingdom, e-mail:
obl20@cam.ac.uk}~~~~~\vspace{0.2cm} \\
\emph{University of Cambridge}}
\maketitle

\begin{abstract}
We consider a class of nonparametric time series regression models in which
the regressor takes values in a sequence space. Technical challenges that
hampered theoretical advances in these models include the lack of associated
Lebesgue density and difficulties with regard to the choice of dependence
structure in the autoregressive framework. We propose an
infinite-dimensional Nadaraya-Watson type estimator, and investigate its
asymptotic properties in detail under both static regressive and
autoregressive contexts, aiming to answer the open questions left by Linton
and Sancetta (2009). First we show pointwise consistency of the estimator
under a set of mild regularity conditions. Furthermore, the asymptotic
normality of the estimator is established, and then its uniform strong
consistency is shown over a compact set of logarithmically increasing
dimension with respect to $\alpha $-mixing and near epoch dependent (NED)
samples. We specify the explicit rates of convergence in terms of the Lambert W
function, and show that the optimal rate is of logarithmic
order, confirming the existence of the curse of infinite dimensionality. 
\newline


\end{abstract}


\section{Introduction}

Nonparametric modelling is a common method for analyzing time series. A
major advantage of this approach is that the relationship between the
explanatory variables under study, denoted by $X=(X_{1},\ldots
,X_{d})^{^{\intercal }}$, and the response, say $Y$, can be modelled without
assuming any restrictive parametric or linear structure; see for example H%
\"{a}rdle (1990), Bosq (1996), or Fan and Yao (2003) for a comprehensive
review. The cost of allowing for this extended flexibility is known as the 
\emph{curse of dimensionality}; Stone (1980, 1982) showed that given a fixed
measure of smoothness $\beta $ allowed on the regression function the best
achievable convergence rate (in minimax sense) $n^{-\beta /(2\beta +d)}$
deteriorates dramatically as the dimension/order $d$ increases. 

In a time series context it is often reasonable and advantageous to model
the dependence upon the infinite past. For example, the $\text{AR}(d)$ model
with $d=\infty $ naturally extends those classical linear models, and
enables the influence of \emph{all past information} to be taken into
account in the modelling procedure, thereby allowing for maximal flexibility
in making statistical inference. It can also be very useful for several
semiparametric applications, and for testing the martingale hypothesis or
the efficient market hypothesis in economics, where the conditional mean $%
E(Y_{t}|\mathcal{F}_{t-1})$ is the object of main interest. By employing
suitable bandwidth adjustments, potential summability issues from the bias
of the estimator can be resolved, and also, the influence of distant
covariates can be suitably downweighted. Not restricting the number of
conditioning variables also has an advantage of avoiding the statistician's 
\emph{a priori} choice of the order $d$ based on some order determination
principles whose validity is subject to be questioned in practical
situations. For these reasons, we are motivated to study a class of
nonparametric time series regression models of infinite order that covers
both static regression and autoregression cases. 

Pagan and Ullah (1988) proposed studying the case where $d\rightarrow
\infty $ in the context of an econometric analysis of risk models. Doukhan
and Wintenberger (2008) studied the autoregressive model with $d=\infty $
under their notion of weak dependence, and showed the existence of a
stationary solution, see also Wu (2011). Linton and Sancetta (2009) tackled the estimation
problem in the context of an autoregressive model and established uniform
almost sure consistency with respect to stationary ergodic sample
observations. There is a vast literature on statistical research on
functional data; typical examples include curves and images, both of which
are infinite-dimensional in nature. Masry (2005) provided a
rigorous treatment of nonparametric regression with dependent functional
data in which $X$ lies in a semi-metric space. Further, Mas (2012) derived the minimax rate of convergence for nonparametric estimation of the regression function on strictly independent and identically distributed regressors. Ferraty and Vieu (2006)
detailed a number of extensions and overview of nonparametric approaches in functional statistics literature.  Geenens (2011) gave an updated accessible summary on nonparametric functional regression, and introduced the term ``curse of infinite dimensionality'', which reflects the difficulties that are expected in nonparametric estimation of infinite dimensional objects due to extreme sparsity. We discuss in the next section the difference between the
functional data framework and our discrete time framework.

Whereas nonparametric regression problems for vectorial regressor have been
exhaustively studied in the literature, statistical theories for their
infinite-dimensional extensions have not been fully established due to
inevitable technical challenges that hindered their widespread popularity.
An obvious difficulty stems from the fact that the usual notion of density $%
p(\cdot )$ does not exist; since there is no $\sigma $-\emph{finite}
Lebesgue measure in infinite-dimensional spaces, the
Lebesgue density (with respect to the infinite product of probability
measures) cannot be defined via the Radon-Nikodym theorem. Consequently,
standard asymptotic arguments for kernel estimators are no longer valid,
notably Bochner's lemma: 
\begin{eqnarray}
\frac{1}{h^{d}}E\left[ \mathcal{K}^{j}\left( \frac{x-X}{h}\right) \right] 
&=&\int \mathcal{K}^{j}(u)p(x-uh)~du~  \notag \\
&\rightarrow &p(x)\Vert \mathcal{K}\Vert _{j}^{j}\quad \text{as}\quad
h\rightarrow 0\label{q2}
\end{eqnarray}%
(under suitable regularity conditions; $j=1,2$), and classical limiting
theories in nonparametric literature cannot be readily extended.

In this paper, we consider an infinite dimensional analogue of the classical
nonparametric regression approach. We propose a local constant type
estimator and investigate its asymptotic properties. In particular, we show
both pointwise and uniform consistency of the estimator and establish its
asymptotic normality under both static regression and autoregressive
contexts with respect to $\alpha $-mixing and near epoch dependent (NED)
samples, respectively. Under some special conditions on the bandwidth, the
explicit rate of convergence is given by specifying the small deviation
probabilities, and the existence of the \emph{curse of
infinite dimensionality} is confirmed. The cases of finite or increasing dimensional
regressors are covered as special cases of our framework.

\section{Some Preliminaries}

Consider the following regression model: 
\begin{equation}
Y=m(X)+\varepsilon  \label{model}
\end{equation}%
where the regressor $X=(X_{1},X_{2},.\ldots )^{^{\intercal }}$ is a random
element taking values in some sequence space $S$, the response $Y$ is a
real-valued variable, and the stochastic error $\varepsilon $ is such that $%
E(\varepsilon |X)=0$ a.s. The objective is to estimate the Borel function $%
m(\cdot )=E(Y|X=\cdot )$ based on $n$ random samples observed from a
strictly stationary data generating process $\{(Y_{t},X_{t})\in \mathbb{R}%
\times S\}_{t\in \mathbb{Z}^{+}}$ having some pre-specified dependence
structure (see section 2.1 below).

This framework is related to the recent advances in statistical research on
functional data, see Ramsey and Silverman (2002) for
general introduction and statistical applications. Recently, successful
attempts have been made to adopt the notion of nonparametric inference on
functional statistics literature, and fundamental theories have been
well-established; Ferraty and Romain (2010) gives a comprehensive review. A
major issue in this field of research lies in extending the statistical
theories of $\mathbb{R}^{d}$ to those of function spaces. In this
literature, the attention is usually on smooth functions that are
approximated and reconstructed from finely discretised grids on some compact
interval. In contrast, the setup in our model (\ref{model}) can be viewed as
looking at an infinite number of discrete observations. Such a difference is
reflected by the fact that the observed data is taken to be a discrete
process $X=(X_{s})$ with unbounded $s\in \mathbb{Z}^{+}$ so that $S=\{f|f:%
\mathbb{N}\rightarrow \mathbb{R}\}$, rather than $X=(X(s))$ with $s\in
\lbrack 0,T]^{k}$ so that $S=\{f|f:[0,T]^{k}\subset \mathbb{R}%
^{k}\rightarrow \mathbb{R}\}$, e.g. curves if $k=1$, images if $k\geq 2$.
Despite isomorphism between separable Hilbert spaces, the discrete nature of
our setting has several fundamental distinctive features that allow us to
look further into many practical applications, a notable example of which is
dependent autoregressive modelling. 

An immediate consequence of the framework is that the tuning parameter can
be imposed for each dimension, allowing one to control the marginal
influence of the regressors. For instance when it is sensible to postulate
that the influence of distant covariates is getting monotonically
downweighted, one may set the marginal bandwidths to increase in lags so as
to impose higher amount of smoothing. Depending on the nature of the
regressor, $S$ may be taken as the space of all infinite real sequences $%
\mathbb{R}^{\mathbb{N}}:=\prod_{j=1}^{\infty }\mathbb{R}_{j}$ formed by
taking Cartesian products of the reals, or various types of its linear
subspaces such as $\ell _{\infty }$, $\ell _{p}$, $c$. We propose to take $S=%
\mathbb{R}^{\mathbb{N}}$ so as to refrain from imposing any prior
restrictions with regard to the choice of the regressor; for example, taking 
$S$ to be the space of bounded sequences excludes the possibility of the
regressors having infinite supports (e.g. Gaussian process). 

\subsection{Dependence structure and leading examples}

A distinctive characteristic of time series data is temporal dependence
between observations. As in the usual multivariate framework, we need some
suitable assumptions on the dependence structure between the samples in
order to derive asymptotic theories and to obtain convergence rates of the
estimator. In nonparametric time series literature, Rosenblatt (1956)'s $%
\alpha $-mixing has been the \emph{de facto} standard choice due to being
the weakest among the mixing-type asymptotic independence conditions. To
name a few earlier works, pointwise and uniform consistency of the local
constant estimator were shown by Roussas (1990) and Andrews (1995),
respectively, and the asymptotic normality was established by Fan and Masry
(1992). Mixing conditions have also been widely used in the context of
dependent functional observations, see e.g. Ferraty et al. (2005), Masry
(2005), and Delsol (2009).\\

\noindent \textsc{Definition 1.} \emph{A stochastic process }$%
\{Z_{t}\}_{t=1}^{\infty }$ \emph{defined on some probability space }$(\Omega
,\mathcal{F},P)$\emph{\ is called} $\alpha $\emph{-mixing or strongly mixing
(cf. `jointly' $\alpha $-mixing if $Z_{t}$ is $\mathbb{R}^{d>1}$-valued)}%
\emph{\ if } 
\begin{equation*}
\alpha (r):=\sup_{A\in \mathcal{F}_{-\infty }^{q},B\in \mathcal{F}%
_{r+q}^{\infty }}|P(A\cap B)-P(A)P(B)|
\end{equation*}%
\emph{is asymptotically zero as }$r\rightarrow \infty $\emph{, where} $%
\mathcal{F}_{a}^{b}$\emph{\ is the }$\sigma $\emph{-algebra generated by }$%
\{Z_{t};a\leq t\leq b\}$\emph{. In particular, we say the process is
algebraically }$($\emph{resp. exponentially}$)$ $\alpha $\emph{-mixing if
there exists some }$c,k>0$ \emph{such that }$\alpha (k)\leq cr^{-k}$ $($%
\emph{resp. if there exists some }$c, \varsigma >0$ \emph{such that }$\alpha (r)\leq
ce^{-\varsigma r}$$)$.\newline

The popularity of the mixing condition stems from the fact that its
associated probability theories have been extensively developed, allowing
simpler theoretical derivations, see for instance Doukhan (1994) or Rio
(2000) for a comprehensive survey. However, several drawbacks have been
pointed out in the literature. First, it is a rather strong technical
condition that is hard to verify in practice. Moreover, it is well known
that even some basic processes are not mixing. e.g. AR(1) with
Bernoulli innovations. 

The primary limitation of mixing conditions that we should be aware of in
our paper arises from the choice of the framework upon which our theories
are developed. Recall that given the sample observations $%
\{Y_{t},X_{t}\}_{t=1}^{n}$ the object of estimation is the conditional mean $%
E(Y_{t}|\mathcal{F})$ where $\mathcal{F}$ represents some information set
determined by the nature of the conditioning variables. There are two
leading cases: the first case is the static regression where the information
set is taken to mean $\sigma (X_{jt};j=1,2,...)$, the $\sigma $-algebra
generated by marginal regressors, for example exogenous variables in
econometric applications. The second case is the autoregression, where $%
X_{tj}=Y_{t-j}$, in which case $\mathcal{F}=\mathcal{F}_{t-1}$ represents $%
\sigma (Y_{s};s\leq t-1)$, the $\sigma $-algebra generated by the sequence
of lags of the responses $(Y_{s})_{s\leq t-1}$. In practice, both types of
conditioning information may be taken into account simultaneously.

In the static regression case the usual joint $\alpha $-mixing condition can
be imposed on the sample data $\{Y_{t},X_{t}\}$ as in the multivariate
framework; since marginal regressors are observed at the same time $t$: $%
X_{t}=(X_{1t},X_{2t},\ldots )^{\intercal }$, assuming joint dependence does
not require additional adjustments. Further, it will be shown later that in
some cases asymptotic properties of the estimator can be invariant to the
cross-dependence structure of $X$. Joint mixing implies both marginal
component processes and any measurable function thereof are mixing\footnote{%
The converse is not necessarily true unless the marginal processes are
independent to others, see Bradley (2005).} (with at most the same
mixing rate $\alpha (\cdot )$). However in the autoregression setting, since
the regressors are taken to be the lags of the response variable, measurable
functions of $X$ depend on the infinite past of the response, i.e. $\Psi
(Y_{t-1},Y_{t-2},\ldots )$ and hence are \emph{not} mixing in general%
\footnote{%
Except for some very special cases; Davidson (1994, Theorem 14.9) gives a
set of technical conditions under which a process with infinite (linear)
temporal dependence is $\alpha $-mixing.}. Due to this reason an alternative
set of dependence conditions is necessary to establish asymptotic theories for the
second framework; we shall adopt the notion of near epoch dependence for the
autoregressive setting and deal with two leading cases separately.\\

\noindent \textsc{Definition 2}\emph{. A stochastic process }$%
\{Z_{t}\}_{t=1}^{\infty }$ \emph{defined on some probability space }$(\Omega
,\mathcal{F},P)$\emph{\ is called near-epoch dependent or stable (in }$L_{2}$%
\emph{-norm) on a strictly stationary} $\alpha $\emph{-mixing process }$%
\{\eta _{t}\}$\emph{\ if the stability coefficients }$%
v_2(m):=E|Z_{t}-Z_{t,(m)}|^{2}$\emph{\ is asymptotically zero as }$%
m\rightarrow \infty $\emph{, where} $Z_{t,(m)}=\Psi_m(\eta_t,\ldots,\eta_{t-m+1})$ for some Borel function $\Psi_m:\mathbb{R}^{m}\rightarrow\mathbb{R}$.\\

A process that is \textquotedblleft near epoch\textquotedblright\
dependent on a mixing sequence is influenced primarily by the
\textquotedblleft recent past\textquotedblright\ of the mixing process and
hence asymptotically resembles its dependence structure; see Lu (2001) for a detailed review. Following the usual convention, e.g. Bierens (1983), we shall take $\Psi_m(\eta_t,\ldots,\eta_{t-m+1})=E(Z_t|\eta_t,\ldots,\eta_{t-m+1})$. We will show that under suitable conditions similar
asymptotic theories can be derived for both regression and autoregression
cases. 

\subsection{Local Weighting}

In order to extend the standard multivariate nonparametric theories to
infinite-dimension and to conduct nonparametric inference on the object of
interest $m(\cdot )$, we first fix the notion of (i) local weighting and
(ii) the measure of closeness between the objects. In the finite dimensional
case, these are respectively done by the (multivariate) kernels $\mathcal{K}%
:(\mathbb{R}^{d},\Vert .\Vert _{2})\rightarrow \lbrack 0,\infty )$ which
control the way the weights are given based on the distance, and the
bandwidth $h$ that regulates the proximity allowed between two objects in
the smoothing procedure. The function $\mathcal{K}$ is called the product
kernel or the spherical kernel depending on the way it is constructed from
the univariate kernel $K(\cdot )$, i.e. $\mathcal{K}^{P}(u)=%
\prod_{j}^{d}K_{j}(u_{j})$ or $\mathcal{K}^{S}(u)=K(\Vert u\Vert _{2})$.

To account for the infinite-dimensional nature of our framework, we extend
the latter approach to construct the weighting functions. We let 
\begin{equation}
\mathcal{K}(u):=K(\Vert u\Vert )  \label{sp}
\end{equation}%
where $u$ is an element of a separable Banach space $(S^{\ast },\Vert .\Vert
)$, and the univariate kernel $K$ is a density function with positive
support, for instance $K(\cdot )=\sqrt{2/\pi }\exp {(-\cdot ^{2}/2)}$. We
shall group the kernels into three subcategories depending on the way how
they are generated. The first two, referred to as Type-I and Type-II kernels
in Ferraty and Vieu (2006) generalize the usual `window' kernels and
monotonically decreasing kernels in finite dimension, respectively. Both
types of kernels are continuous on compact support $[0,\lambda ]$.\newline

\noindent\textsc{Definition 3}\emph{. A function }$K:[0,\infty)\rightarrow
[0,\infty)$\emph{\ is called a kernel of type}$-I$\emph{\ if it integrates
to 1, and if there exist strictly positive real constants }$C_{1},C_{2}$%
\emph{\ (with }$C_{1}<C_{2}$\emph{) for which } 
\begin{equation}
C_{1}\emph{1}_{[0,\lambda]}(u)\leq K(u)\leq C_{2}\emph{1}_{[0,\lambda ]}(u),
\label{q3}
\end{equation}
\emph{where }$\lambda$\emph{\ is some fixed positive real number. Also, a
function }$K:[0,\infty)\rightarrow [0,\infty)$\emph{\ is called a kernel of
type}$-II$\emph{\ if it satisfies (\ref{q3}) with }$C_{1}\equiv0 $\emph{,
and is continuous on $[0,\lambda]$ and differentiable on $(0,\lambda)$ with
derivative }$K^{\prime}$\emph{\ that satisfies } 
\begin{equation*}
C_{3}\leq K^{\prime}(u)\leq C_{4}
\end{equation*}
\emph{for some real constants }$C_{3},C_{4}$\emph{\ such that }$-\infty
<C_{3}<C_{4}<0 $.\newline

The definition above suggests that the uniform kernel on $[0,\lambda ]$ is a
type-I kernel, and the Epanechnikov, Biweight and Bartlett kernels belong to
the class of Type-II kernels. Those with semi-infinite support, for example
(one-sided) Gaussian, are covered by the last group, which we shall call the
Type-III kernels. \newline

\noindent \textsc{Definition 4}\emph{. A function }$K:[0,\infty )\rightarrow
\lbrack 0,\infty )$\emph{\ is a kernel of type}$-III$\emph{\ if it
integrates to 1, and if it is of exponential type; that is, }$K(r)\propto
\exp (Cr^{\beta })$ \emph{for some constants }$\beta $\emph{\ and }$C$. 

\subsection{Small deviation}

The \emph{small ball (or small deviation) probability } plays a crucial role
in establishing the asymptotic theories in this paper. Let $S^{\ast }$ be a
separable sequence space that is complete with respect to some norm $\Vert
.\Vert $; then the small ball probability of an $S^{\ast }$-valued random
element $Z$ is a function defined as%
\begin{equation}
\varphi _{z}(h):=P\left( \Vert z-Z\Vert \leq h\right) .  \label{q5}
\end{equation}%
The probability is called centered if $z=0$, and shifted (with respect to a
fixed point $z\in S^{\ast }$) if otherwise. The relation between the two
quantities cannot be explicitly specified in general, and will be expressed
in terms of the Radon-Nikodym derivative (See Assumption B1 below).

The name \emph{small ball} stems from the fact that we are interested in its
asymptotic behaviour as $h$, the bandwidth sequence in the context of
nonparametric estimation, tends to zero. The function can be thought of as a
measure for how much the observations are densely \emph{packed} or \emph{%
concentrated} around the fixed point $z$ with respect to the associated norm
and the reference distance $h$. From the definition it is straightforward to
see that $\varphi(h)\rightarrow0$ as $h\rightarrow 0$, and that $n\varphi(h)$
is an (approximate) count of the number of observations whose influence is
taken into account in the smoothing procedure. When $Z$ is a $d$-dimensional
continuous random vector with Lebesgue density $p(\cdot)>0$, it can be
readily shown that the shifted small ball (with respect to the usual
Euclidean norm) is given by 
\begin{equation*}
\varphi _{z}(h)=V_{d}h^{d}p(z)\propto h^{d},
\end{equation*}
where $V_{d}=\pi^{d/2}/\Gamma(d/2+1)$ is the volume of $d$-dimensional unit
sphere.

However when $Z$ takes values in an infinite-dimensional space, it is
generally difficult to specify the exact form of the small ball probability,
and its behaviour varies depending heavily on the nature of the associated
space and its topological structure. Due to non-equivalence of norms, it is
intuitively clear that the \emph{speed} at which $\varphi(h)$ converges to
zero is affected by the choice of $\|.\|$. In any case, a rapid decay is
expected due to extreme sparsity of data in infinite dimensional spaces.

One possible example of $S^{\ast }$ includes $(\ell _{r},\Vert .\Vert _{r})$%
, the space of $r$-th power summable sequence equipped with the $\ell _{r}$
norm. As we shall reiterate below in the main text, we will take our main
focus to be on the case of $r=2$, since the behaviour of the small ball has
been most studied in the case of square summable random variables. The case
of $r>2$ may also be derived without much difficulty, provided that the
moment conditions are modified appropriately.

Writing the expected value of the kernel in terms of the small ball
probability 
\begin{align}
E\mathcal{K}\bigg(\frac{z-Z}{h}\bigg)& =EK\bigg(\frac{\Vert z-Z\Vert }{h}%
\bigg)  \notag \\
& =\int K(u)~dP_{\Vert z-Z\Vert /h}(u)=\int K(u)~d\varphi _{z}(uh),
\label{q6}
\end{align}%
we are able to bypass the difficulties mentioned in the introduction,
and establish the convergence of the integrals without explicitly
requiring the existence of the density. \newline

\noindent \textsc{Lemma 1}. Ferraty and Vieu (2006, Lemma 4.4).\emph{\ If }$%
K $ \emph{is a type-I kernel, then it satisfies } 
\begin{equation}
C_{1}^{j}\leq \frac{1}{\varphi _{z}(h\lambda )}\int_{0}^{\lambda
}K^{j}(v)~d\varphi _{z}(vh)\leq C_{2}^{j},~~j=1,2  \label{q7}
\end{equation}%
\emph{where }$C_{1},C_{2}$\emph{\ are strictly positive real constants
defined in Definition 3. Furthermore, if the kernel }$K$ \emph{is type-II
and if} \emph{there exists some }$\varepsilon _{0}>0$\emph{\ and a constant }%
$C_{5}>0$\emph{\ such that }$\forall \varepsilon <\varepsilon
_{0},~\int_{0}^{\varepsilon }\varphi _{x}(u)du>C_{5}\varepsilon \varphi
_{x}(\varepsilon )$, \emph{then we have} 
\begin{equation}
C_{6}^{j}\leq \frac{1}{\varphi _{z}(h\lambda )}\int_{0}^{\lambda
}K^{j}(v)~d\varphi _{z}(vh)\leq C_{7}^{j},~~j=1,2  \label{q8}
\end{equation}%
\emph{where the constants }$C_{6}=-C_{5}C_{4}$\emph{\ and }$C_{7}=\sup_{s\in
\lbrack 0,\lambda ]}K(s)$\emph{\ are strictly positive}.
\newpage

Since Lemma 1 holds for every $h>0$, it readily follows by the squeeze
theorem that the following result holds for any kernels of type-I and II:%
\newline

\noindent\textsc{Corollary 1}.\emph{\ If the kernel $K$ is either type-I or
type-II, then for $j=1,2$ we have} 
\begin{equation}
\frac{1}{\varphi_{z}(h\lambda)}E\left[\mathcal{K}^j\left( \frac{z-Z}{h}%
\right)\right] \longrightarrow\xi_{j}\quad\text{as}\quad h\rightarrow0^{+} ,
\label{q10}
\end{equation}
\emph{where $\xi_1$ and $\xi_2$ are some strictly positive real constants.}%
\newline

This result can be seen as an infinite-dimensional analogue of Bochner's
lemma (\ref{q2}): i.e. for $Z\in \mathbb{R}^{d}$, $h^{-d}E\mathcal{K}%
((z-Z)/h)\rightarrow p(z)>0$, and will play a fundamental role in
constructing asymptotic theories of our estimator. It is obvious that $\xi
_{j}$ is bounded below and above by $C_{1}^{j}$ and $C_{2}^{j}$,
respectively (or $C_{6}^{j}$ and $C_{7}^{j}$ depending on the choice of the
kernel). With specific choices of kernels and regressors the exact values of
the constants may be specified.\newline

\textsc{Remark}. A natural question one may ask is whether (\ref{q10}) would
still be valid for other kernels, in particular those with semi-infinite
support, namely the Type-III kernels. In finite $d$-dimensional frameworks,
it is well known that a set of standard assumptions including $\Vert u\Vert
^{d}K(u)\rightarrow 0$ as $u\rightarrow \infty $ is sufficient for showing (%
\ref{q2}), see for instance Pagan and Ullah (1999, Lemma 1). However, in
the infinite-dimensional setting the answer is negative in most usual cases
where the kernel is of exponential type (e.g. Gaussian kernel). Whereas the
lower bound of the limit can be easily constructed via Chebyshev's
inequality: 
\begin{equation*}
(0<)\exp (-c_{\delta }\delta )\leq \lbrack P(V\leq \delta )]^{-1}E\exp
(-c_{\delta }V),
\end{equation*}%
the upper bound may not exist, and the rate at which the small ball
probability decays to zero may dominate the speed at which the integral (\ref%
{q6}) converges to zero. This claim cannot be formally verified for all
general cases because (as mentioned above) there is no unified result for
the asymptotic behaviour of small deviation available. Nonetheless, the idea
can be sketched in the common case where the asymptotics of the distribution
function (i.e. small deviation) is of exponential order: $P(V\leq \delta
)\sim \exp (-C{\delta }^{-{\theta }/({\theta }+1)})$ as $h\rightarrow 0$ for
some constants $C$ and $\theta $; by de Bruijn's exponential Tauberian
theorem (see Bingham et al. (1987)) the necessary and
sufficient condition for such a case is the following limiting behaviour of
the Laplace transform near infinity: 
\begin{equation*}
E[\exp (-c_{\delta }V)]\sim \exp \big(C^{\prime }\cdot c_{\delta }^{-\theta }%
\big)\quad \text{as}\quad c_{\delta }\rightarrow \infty 
\end{equation*}%
for some constant $C^{\prime }$. (e.g. Take, for the example of Gaussian
kernel, $V=\Vert z-Z\Vert ^{2}$, $\delta =h^{2}$, $c_{\delta }=2^{-1}h^{-2}$%
) Difference in the order of convergence implies divergence of the limit (%
\ref{q10}). For this reason, we shall confine our attention to compactly
supported kernels hereafter.

\subsection{The Estimator}

Consider the regression problem with $\mathbb{R}^{\mathbb{N}}$-dimensional
regressor $X=(X_{1},X_{2},\ldots )^{^{\intercal }}$, where the regression
operator is nonparametrically estimated with an estimator associated with a
bandwidth matrix $H:=diag(\underline{h})=diag(h_{1},h_{2},...)$. No further
condition is explicitly imposed in the first instance except that the
bandwidth sequence is appropriately chosen in such a way that a norm $\Vert
.\Vert $ can be admitted to the \emph{weighted regressor}. As an
expositional example, let us assume $h_{j}=\phi _{j}h$ where $\{\phi
_{j}^{-1}\}_{j}$ is some square-summable sequence (e.g. $\phi _{j}=j^{p}$,
for some $p\geq 1$).

Then by Kolmogorov's three-series theorem, the sequence of weighted
regressor $\phi _{j}^{-1}X_{j}$ is square summable, w.p.1., provided that $%
X_{j}^{\prime }s$ are independent with finite variance and satisfy 
\begin{equation*}
\sum_{j=0}^{\infty }E\min \big\{1,\phi _{j}^{-2}X_{j}^{2}\big\}<\infty .
\end{equation*}%
so that $(\phi _{1}^{-1}X_{1},\phi _{2}^{-1}X_{2},...)^{^{\intercal }}=:Z$
is $(\ell _{2},\Vert .\Vert _{2})$-valued. Since this is a mild set of
requirements, we see that sufficient flexibility is given while
sophisticated mathematical developments are allowed. In the autoregressive
context, the weighting sequence can be chosen to be non-decreasing so that
the \textquotedblleft relative influence\textquotedblright\ of the marginal
regressor decreases in index (lag).

Seeing the arguments sketched above we shall assume from now on that $Z$ is
normed with $\Vert .\Vert =\Vert .\Vert _{2}$, and (with an abuse of
notation) extend the usual definition of (shifted) small ball probability to
account for varying bandwidths $h_{j}$ and generalized support $[0,\lambda ]$
as follows: 
\begin{equation}
\varphi _{x}(\underline{h}\lambda ):=~P\big(\Vert H^{-1}(x-X)\Vert \leq
\lambda \big),\label{q11}
\end{equation}%
or equivalently $P(X\in \mathcal{E}(x,\underline{h}\lambda ))$, where $%
\mathcal{E}$ is the infinite-dimensional hyperellipsoid centred at $x\in 
\mathbb{R}^{\mathbb{N}}$, and $\lambda $ is the constant defined in section
2.2. Clearly, we have $\varphi _{x}(\underline{h}\lambda )=\varphi
_{z}(h\lambda )$. For later reference, we also define the joint small ball
probability as the joint distribution 
\begin{equation}
\psi _{x}(\underline{h}\lambda ;i,j):=~P\big((X_{i},X_{j})\in \mathcal{E}%
(x,\lambda \underline{h}\big)\times \mathcal{E}(x,\lambda \underline{h})\big)%
.  \label{smallball}
\end{equation}

We are finally in a position to introduce our estimator. We propose to
estimate $m(x)=E(Y|X=x)$,~$x\in \mathbb{R}^{\mathbb{N}}$ with the following
local constant type estimator: 
\begin{equation}
\widehat{m}(x):=\frac{\sum_{t=1}^{n}\mathcal{K}\Big(H^{-1}(x-X_{t})\Big)Y_{t}%
}{\sum_{t=1}^{n}\mathcal{K}\Big(H^{-1}(x-X_{t})\Big)}\equiv \frac{%
\sum_{t=1}^{n}K\Big(\Vert H^{-1}(x-X_{t})\Vert \Big)Y_{t}}{\sum_{t=1}^{n}K%
\Big(\Vert H^{-1}(x-X_{t})\Vert \Big)},  \label{q12}
\end{equation}%
defined with respect to $n$-sample time series observations $%
\{Y_{t},X_{t}\}_{t=1}^{n}$. 

The estimator can be viewed as an infinite-dimensional generalization of the
standard multivariate Nadaraya-Watson estimator, and is a special case of
the one considered in Ferraty and Vieu (2002), Masry (2005) and references
therein for general functional data. The aim of this paper is to examine
some key asymptotic properties of the estimator and establish its
statistical theory. 

\section{Asymptotic Properties}

In this section we introduce the main results of our paper, deriving some
large sample asymptotics of the proposed estimator (\ref{q12}). Consistency
is shown in both pointwise and uniform sense and the asymptotic normality is
established. Rates of convergence are specified under sets of regularity
conditions. All proofs are detailed in the appendix.

We shall consider two different cases: 1) static regression and 2)
autoregression. We therefore assume one of the two different sets of
conditions on the dependence structure of the data generating process
specified below. Assumption S1 below concerns with the static regression
case where we have exogenous regressors that are jointly observed in time in
a mildly dependent manner. No restriction is needed as regards
cross-sectional dependence between the regressor, although necessary
conditions will be imposed when needed at the later stage (see Assumptions D
below). The second assumption concerns with the case where we observe a $%
\mathbb{R}$-valued response that is $\alpha $-mixing in time. Due to the
reasons sketched in section 2.1, we adopt the notion of near epoch
dependence to describe the dependence structure of infinite functions of the
response variables. From the assumption we see that there is a trade-off
between the degree of mixing rate and the possible order of moments allowed
for the response variable $2+\delta $.\\

\textsc{Assumptions S}

\begin{enumerate}
\item[S1.] \emph{The marginal regressors }$X_{1t},X_{2t},X_{3t},\ldots $%
\emph{\ are exogenous variables, and the sample observations }$%
\{Y_{t},X_{t}\}_{t=1}^{n}=\{Y_{t},(X_{1t},X_{2t},\ldots )\}_{t=1}^{n}$\emph{%
\ are (jointly) arithmetically }$\alpha $\emph{-mixing with rate }$%
k>(2\delta +4)/\delta $\emph{, where $\delta $ is as defined in Assumption
A4 below}.

\item[S2.] \emph{The regressors are lags of the response, i.e. }$%
X_{it}=Y_{t-i}~\forall i$\emph{, and the process }$\{Y_{t}\}_{t=1}^{n}$\emph{%
\ is arithmetically }$\alpha $\emph{-mixing with }$k>(2\delta +4)/\delta $. 
\emph{Also, the process }$K_{t}:=K(\Vert H^{-1}(x-X_{t})\Vert )$%
\emph{\ is near epoch dependent on }$Y_{t}$\emph{, and there exists some $m=m_n\rightarrow\infty$ such that the rate of
stability for $K_t$ denoted }$v(m_{n})=v(m)$\emph{\ satisfies} 
\begin{equation}
~v(m)^{1/2}[\varphi _{x}(\underline{h}\lambda )]^{-(2\delta +3)/(2\delta
+2)-1}n^{1/(2(\delta +1))}\rightarrow 0\qquad as~~n\rightarrow \infty .
\label{f2a}
\end{equation}
\end{enumerate}

\subsection{Pointwise consistency}
Pointwise consistency of the local constant estimator was first studied by
Watson (1964) and Nadaraya (1964) with respect to i.i.d sample for the case
of univariate regressor, i.e. $d=1$. Their result was extended to the
multivariate case by Greblicki and Krzyzak (1980) and Devroye (1981).
Robinson (1983) and Bierens (1983) were amongst the earliest papers that
worked on consistency of the estimator with respect to dependent data (both
static regression and autoregression were allowed in their frameworks),
followed by Roussas (1989), Fan (1990), and Phillips and Park (1998) to name
a few out of numerous papers. The case of functional regressor was first
studied by Ferraty and Vieu (2002), and references therein under various
sets of regularity conditions.

In this section we show how these results can be extended specifically to
our framework, proving pointwise weak consistency of the estimator (\ref{q12}%
) with respect to dependent sample satisfying either S1 or S2. A set of
assumptions required for the theory is now introduced, and the series of
arguments that leads to the conclusion is sketched.\newline

\noindent\textsc{Assumptions A}

\begin{enumerate}
\item[A1.] \emph{The regression operator }$m:\mathbb{R}^{\mathbb{N}%
}\rightarrow \mathbb{R}$\emph{\ is continuous in some neighbourhood of }$x$

\item[A2.] \emph{The marginal bandwidths satisfy }$h_{j}=h_{j,n}\rightarrow0$%
\emph{\ as }$n\rightarrow\infty$\emph{\ for all} $j=1,2,\ldots$\emph{, where}
$diag(h_{1},h_{2},...)=diag( \underline{h})=H$\emph{\ is the bandwidth
matrix, and the small ball probability obeys }$n\varphi_{x}(\underline{h}%
\lambda )\rightarrow\infty$\emph{, where }$\varphi_{x}(\underline{h}\lambda
):=P(\|H^{-1}(x-X)\|\leq\lambda)\rightarrow0$\emph{\ as }$n\rightarrow\infty$%
.

\item[A3.] \emph{The kernel }$K$\emph{\ is either type}$-I$\emph{\ or type}$%
-II.$

\item[A4.] \emph{The response }$Y_{t}$\emph{\ satisfies }$E\big(|Y_{t}|^{2+\delta }\big)%
\leq C<\infty $ for some $C,\delta >0.$

\item[A5.] \emph{The joint small ball probability (\ref{smallball})
satisfies }$\psi _{x}(\underline{h}\lambda ;i,j)=O(\varphi _{x}(\lambda 
\underline{h})^{2}),~\forall i\neq j$.

\item[A6.] \emph{The conditional expectation }$E\big(|Y_tY_s||(X_t,X_s)\big)\leq C<\infty$\emph{ for all }$t, s$.
\end{enumerate}

\textsc{Remark.} Asymptotic unbiasedness of the estimator requires the
continuity assumption A1; in other words, the estimator is unbiased at every
point of continuity. Upon imposing further smoothness condition on the
regression operator, the rate of convergence for the bias term can be
specified (as we shall do later). Assumption A2 can be thought of as an
extension of the usual bandwidth conditions that are assumed in
finite-dimensional nonparametric literature. As discussed before, $n\varphi(%
\underline{h}\lambda)$ can be understood as an approximate number of
observations that are ``close enough'' to $x$. 
Therefore, it is sensible to postulate that $n\varphi(\underline{h}%
\lambda)\rightarrow\infty$ as $n\rightarrow\infty$, meaning that the point $%
x $ is visited many times by the sample of data as the size of sample grows
to infinity. This is in line with the usual assumption $nh^{d}\rightarrow%
\infty$ when $X\in\mathbb{R}^{d}$, in which case the small ball probability
is given by $\varphi(h)\propto h^{d}$. 
Condition A3 suggests that our result is valid under flexible choices of the
compactly supported kernels. The standard moment condition A4 can be
replaced by the boundedness condition $|Y|\leq C$, a.s. to simplify the
proof, e.g. Billingsley's inequality to bound asymptotic covariance terms.
However we shall not consider this possibility here because this costs
exclusion of the distributions with unbounded support. Conditions A5 and A6 are assumed to control the asymptotics of the covariance terms.\newline

To sketch the idea, we write $K_{t}:=K(\|H^{-1}(x-X_{t})\|)$ for the sake of
simplicity of presentation (note its dependence upon $X_{t}$), and express
the estimator (\ref{q12}) as 
\begin{equation}
\widehat{m}(x):=\frac{\sum_{t=1}^{n}K\big(\|H^{-1}(x-X_{t})\|\big)Y_{t}}{%
\sum_{t=1}^{n}K\big(\|H^{-1}(x-X_{t})\|\big)}=\frac{\frac{1}{n}\sum_{t=1}^{n}%
\frac{K_{t}}{EK_{1}}Y_{t}}{\frac{1}{n}\sum_{i=1}^{n}\frac{K_{t}}{EK_{1}}}=%
\frac{\widehat{m}_{2}(x)}{\widehat{m}_{1}(x)}.  \label{fraction}
\end{equation}

\noindent We then employ the following decomposition: 
\begin{align}
\widehat{m}(x)-&m(x)= \frac{\widehat{m}_{2}(x)}{\widehat{m}_{1}(x)}-m(x) = 
\frac {\widehat{m}_{2}(x)-m(x)\widehat{m}_{1}(x)}{\widehat{m}_{1}(x)}  \notag
\\
& = \frac{E\widehat{m}_{2}(x)-m(x)E\widehat{m}_{1}(x)}{\widehat{m}_{1}(x)}+ 
\frac{[\widehat{m}_{2}(x)-E\widehat {m}_{2}(x)]-m(x)[\widehat{m}_{1}(x)-E%
\widehat{m}_{1}(x)]}{\widehat{m}_{1}(x)},  \label{decomp1}
\end{align}
where clearly $E\widehat{m}_{1}(x)=1$. We shall show consistency by proving
that the `bias part' $E\widehat{m}_{2}(x)-m(x)$ and the `variance part' $[%
\widehat{m}_{2}(x)-E\widehat{m}_{2}(x)]-m(x)[\widehat{m}_{1}(x)-1]$ are
negligible in large sample. As for the latter term, it suffices to show mean
squared convergence of $\widehat{m}_{2}(x)-E\widehat{m}_{2}(x)$ to zero
because then $\widehat{m}_{1}(x)\rightarrow^p1$ will readily follow.\newline

\noindent\textsc{Theorem 1.}\emph{\ Suppose A1-A5 hold.} \emph{Then the
estimator (\ref{q12}) with respect to the sample observations} $%
\{Y_{t},X_{t}^{\intercal}\}_{t=1}^{n}$\emph{\ satisfying either S1 or S2 is
weakly consistent for the regression operator }$m(x)$\emph{. That is, as }$%
n\rightarrow\infty$ 
\begin{equation}
\quad\widehat{m}(x)\overset{P}{\longrightarrow}m(x).  \label{q13} \\
\end{equation}

\indent\textsc{Remark.} It is trivial to see that this result holds for
i.i.d. data $\{(Y_{t},X_{t}^{\intercal});~t\in\mathbb{Z}\}$ as well. The
arguments simply becomes less involved because the covariance term does not
need to be considered anymore. An alternative way to approach would be to
apply a suitable exponential inequality, in which case some
additional/different assumptions may be needed.\newline

In the following section, we shall specify the convergence rates and
establish the asymptotic normality by imposing a set of additional
regularity conditions. 

\subsection{Asymptotic Normality}

Rigorous studies on the limiting distribution of the standard
Nadaraya-Watson estimator can be traced back to Schuster (1972) and Bierens
(1987), where the case of univariate and multivariate regressors was
considered, respectively. The case of dependent samples was studied by
Robinson (1983), Bierens (1983) and then by many others under a wide range
of different frameworks. Constructing asymptotic normality given a dependent
sample data typically involves the use of Bernstein's ``big-small'' blocking
argument followed by the standard Lindeberg-Feller central limit theorem.

General distributional theories for Nadaraya-Watson type estimators in a
semi-metric space was established by Masry (2005, Theorem 4) and Delsol
(2009). This paper is different from the existing literature in two
perspectives. First, our specific setting allows autoregressive time series
modelling with respect to dependent sample data. Second, whereas the final
results of the existing papers were given in terms of abstract functions,
our results are presented with the explicit rate of convergence, thereby
allowing feasible applications. 

The primary object of this section is to outline this procedure in detail,
and to introduce the main theories and some interesting consequences
thereof. To reiterate, both the i.i.d. case (in other words, when the
marginal regressors $X_{j}$ are independent and identically distributed) and
also a dependent case shall be allowed. Specifically, we introduce how
independence restriction can possibly be moderated to allow for a mild
dependence structure. In particular, the second assumption below specifies
the extent to which cross-sectional dependence can be allowed on the
marginal regressors in our theory whilst allowing for specification of the
exact form of the convergence rate of the estimator. \newline

\noindent \textsc{Assumptions D.} \emph{The real-valued stochastic process $%
\{X_{s}\}_{s=1}^{\infty }$ satisfies $EX_{s}^{4}\leq C<\infty $, and is
either:}

\begin{enumerate}
\item[D1.] \emph{independent and identically distributed},~or

\item[D2.] \emph{stationary, and admits a causal moving average
representation:} 
\begin{equation}
X_{s}=\sum_{j=0}^{\infty }a_{j}\epsilon _{s-j},  \label{crsec}
\end{equation}%
\emph{where }$a_{j}$\emph{\ is a square summable sequence, and }$\{\epsilon
_{j}\}_{j}$\emph{\ is an independent and identically distributed standard
Gaussian sequence.}\\
\end{enumerate}

\textsc{Remark}. In either case marginal regressors are required to be
identically distributed; an additional distributional assumption is made in
B2 below. Nonetheless, possible degree of dependence allowed in D2 is very
mild and general. (In fact, the condition can even be generalized to require
the existence of a non-causal MA representation, whose necessary and
sufficient condition is simply the existence of the spectral density.) As an
example, if a stationary stochastic process $\{X_{s}\}_{s=1}^{\infty }$ is $%
\alpha $-mixing, then it always has a moving average representation of the
form (\ref{crsec}) provided it is Gaussian. This is because any $\alpha $%
-mixing process is regular\footnote{%
in the sense of e.g. Bulinskii and Shiryaev (2003, p.248)} by definition, so
is linearly regular when it is Gaussian, and hence (with stationarity)
admitting the Wold decomposition with independent Gaussian innovations by 
Corollary 17.3.1 of Ibragimov and Linnik (1971). For this reason we see that
each D1 and D2 is consistent with the case allowed in Assumption S1 and S2,
respectively (because in Assumption S2 the regressors are cross-sectionally
mixing as they are temporal lags of the response which is mixing by
assumption), although the dependence structure specified in D2 can be
allowed also for the static regression case (i.e. S1). This suggests that
there is absolutely no need to assume independence between marginal
regressors in our model (\ref{model}) under Gaussianity. In particular,
convergence rates of our estimator are shown to be invariant (upto some
constant factor) to the choice between D1 and D2. Lastly, the requirement
for finite fourth moment is to ensure that the squared marginal regressors
have finite second moment (see below); obviously, in the autoregressive
framework (Assumption S2), this forces $\delta $ in Assumption A4 to be
greater than or equal to $2$.

We now proceed to introduce some main assumptions needed for distributional
theories. 

\subsubsection{The `bias component'}

The first part concerns with the asymptotic `bias', where Assumptions A is
strengthened by imposing additional smoothness conditions and suitable
bandwidth adjustments. It is a set of sufficient conditions under which one
is able to specify the finite sample properties of the estimator/ the exact
upper bound of the asymptotic bias. Note that alternatively, Fr\'{e}chet
differentiability type condition can be imposed, as was done in Mas (2012).\\

\noindent \textsc{Further Assumptions A}

\begin{enumerate}
\item[A7.] \emph{Further to Assumption A2, the marginal bandwidths satisfy }$%
h_{j}=\phi_j\cdot h$\emph{\ for some positive real number $\phi_j$ where $%
h=h_n\rightarrow0$ as $n\rightarrow\infty$}.

\item[A8.] \emph{The regression operator }$m:\mathbb{R}^{\mathbb{N}%
}\rightarrow\mathbb{R}$\emph{\ satisfies } 
\begin{equation}
\big|m(x)-m(x^{\prime })\big|\leq\sum_{j=1}^{\infty}c_{j}\big|%
x_{j}-x_{j}^{\prime }\big|^{\beta}  \label{adit}
\end{equation}
\emph{for every }$x^{\prime },x=(x_{1},x_{2},\ldots)^{\intercal}\in\mathbb{R}%
^{\mathbb{N}}$\emph{\ and some real constant }$\beta\in(0,1]$\emph{, where} $%
c_{j}$\emph{\ is some sequence of real constants for which }$%
\sum_{j=1}^{\infty}c_j\leq 1$\emph{\ and }$\sum_{j=1}^{\infty}
c_{j}\phi_j^{\beta }<\infty$.\newline
\end{enumerate}

\textsc{Remark.} Roughly speaking, these assumptions are imposed to ensure
the bias to be \textquotedblleft well-bahaved\textquotedblright . Assumption
A7 extends the previous bandwidth condition A2. Obviously, it is consistent
with what was previously assumed in A2 since $h\rightarrow 0$ implies
coordinate-wise convergence of each marginal bandwidths in large sample.
With this condition one is able to write the asymptotic bias expression and
the order of the bias-variance balancing bandwidth in terms of the common
factor $h$. It is possible to dispense with this condition at the cost of
imposing minor modifications in A7; the asymptotic bias will then be written
in terms of the infinite sum of a weighted marginal bandwidth $h_{j}$, whose
convergence needs to be assumed. Furthermore, we remark that if the variance
term is not concerned, the sequence of marginal coefficients $\phi _{j}$
does not necessarily have to be monotonically increasing in $j$, which would
allow more flexible choices of bandwidths. However, we shall assume some
increment condition in Assumptions B below, since such a restriction is
needed for the variance component.

Assumption A8 replaces and strengthens Assumption A1, and can be thought of
as a variant of the usual H\"{o}lder-type smoothness conditions. With the
condition, the regression operator becomes a contraction mapping; it will be
seen in the appendix that summability of the bias is ensured and the order
of convergence rates for the bias term is specified, see also (\ref{bias}).
One example of the contraction constant $c_{j}$ would be $\exp (-j)$.
Indeed, the contribution from each marginal dimension decrease in lag or
index under this assumption.

In the context of autoregression where $X_{j}\equiv Y_{t-j}$ for any $j$, it
is imperative to ensure the existence of stationary solution $\{Y_{t}\}$,
since the model is essentially an infinite number of difference equations
where finiteness of the solution is not automatically guaranteed. In the
study of a class of general nonlinear AR($d$) models, Duflo (1997) and G\"{o}%
tze and Hipp (1994) assumed what is called the Lipschitz mixing condition
(or the strong contraction condition), which is essentially (\ref{adit})
replaced by finite $d$-sum on the right hand side. In our context,
Assumption A6 plays such a role; Doukhan and Wintenberger (2008) showed that
(\ref{adit}) with $\sum_{j=1}^{\infty }c_{j}<1$, is sufficient for the
existence of a stationary solution of the form 
\begin{equation}
Y_{t}=f(\varepsilon _{t},\varepsilon _{t-1},\ldots ),
\end{equation}%
where $f$ is some measurable function. Wu (2011) arrived at the same
conclusion under the assumption of $\sum_{j=1}^{\infty }c_{j}=1$; the
restriction on $c_{j}$ in our assumption reflects the findings of their
papers.

\subsubsection{The `variance component'}

We now move on to the second chunk of assumptions that concerns with the
`variance part'. As before, vectors $Z$ and $z$ are taken to mean $(\phi
_{1}^{-1}X_{1},\phi _{2}^{-1}X_{2},...)^{^{\intercal }}$ and $(\phi
_{1}^{-1}x_{1},\phi _{2}^{-1}x_{2},...)^{^{\intercal }}$, respectively,
where the vector $x=(x_{1},x_{2},\ldots )^{^{\intercal }}$ is the point at which
estimation is made.
\newpage

\noindent \textsc{Assumptions B}

\begin{enumerate}
\item[B1.] \emph{The induced probability measure $P_{z-Z}$ is dominated by
the measure $P_Z$, and its Radon-Nikodym density $dP_{z-Z}/dP_{Z}=:p^{*}$ is
sufficiently smooth in some neighbourhood of zero, and is bounded away from
zero at 0; i.e., $p^{*}(0)>0.$} 

\item[B2.] \emph{Given the regressor vector $X=(X_{1},X_{2},...)^{\intercal}%
\in\mathbb{R}^{\mathbb{N}}$, the distribution $F$ of $X_s^2$ is regularly
varying near zero with strictly positive index.}

\item[B3.] \emph{Further to A5, the bandwidth sequence satisfies $%
h_{j}=j^{p}h$ with $p>1$; i.e. }$\phi_j=j^{p}$.

\item[B4.] \emph{The conditional variance }$\text{Var}[Y_t|X_t=u]=%
\sigma^{2}(u)$\emph{\ is continuous in some neighbourhood of }$x$\emph{;
i.e. }$\sup_{u\in \mathcal{E}(x,\underline{h}\lambda)}[\sigma^{2}(u)-%
\sigma^{2}(x)]=o(1).$\emph{\ Similarly, the cross-conditional moment }$%
E[(Y_t-m(x))(Y_s-m(x)|X_t=u,X_s=v]=\sigma(u,v),~t\neq s$\emph{\ is
continuous in some neighbourhood of }$(x,x)$.\newline
\end{enumerate}

\textsc{Remark.} The first condition concerns with a transition of the shifted
small ball probability to the centred quantity; it is taken from Mas (2012), where detailed discussions can be found. In the Gaussian framework, the
limiting behaviour between the two is well-understood, see Zolotarev (1988)
or de Acosta (1983) for example, unlike the non-Gaussian cases, see Skorohod (1967). Condition B2 is equivalent to saying 
\begin{equation*}
\lim_{x\rightarrow\infty}\frac{F(1/(\gamma x))}{F(1/x)}=\gamma^{\rho}
\end{equation*}
with strictly negative index of variation $\rho<0$. This marks the close
relationship between the laplace transform and the small ball probability,
see Lifshits (1997), Mas (2012) for example for detailed discussions.

We require the function $F(1/x)$ to be regularly varying in order to ensure
that the small ball probability \emph{well-behaves} (near infinity) in
asymptotic sense. Since only those having strictly negative $\rho $ satisfy
the condition, a regressor must have a distribution $F(1/x)$ that decreases
(as $x\rightarrow \infty $) at a \emph{reasonable speed}. By reasonable we
mean that the relative weight of decrease follows a power law, and the
variation should be continuous. A large class of common distributions
satisfies the condition; for example, Gamma, Beta, Pareto, Exponential,
Weibull, and also chi-squared distribution (in which case each $X_{s}$ is
Gaussian). We refer the reader to Conditions $I$, $L$ and relevant
discussions thereof in Dunker et al. (1998) for details. Next, the
increment condition on the bandwidth B3 is assumed so as to specify the
explicit behaviour of the small ball probability. We remark that different
coefficients such as exponential weights (i.e. $h_{j}=e^{j}h$) can also be
allowed, leading to similar distributional results given in the subsequent
sections; however, we shall confine our attention to the case of polynomial
law in this paper for the sake of simplicity of exposition. The standard
conditions in B4 are assumed to deal with the asymptotics of
the variance and covariance terms.\newline

With reference to (\ref{decomp1}) we are now able to derive the following
results for the bias and variance components, in view of Assumptions A7, A8
and B, and Corollary 1: 
\begin{align}
\mathcal{B}_n:=&~\Big[E\widehat{m}_2(x)-m(x)\Big]\leq
h^{\beta}\lambda^{\beta}\sum_{j=1}^{\infty}c_jj^{p\beta}  \label{bias} \\
\mathcal{V}_n:=&~\text{Var}\Big[\widehat{m}_2(x)-E\widehat{m}_2(x)\Big]\simeq%
\frac{\sigma^{2}(x)\xi_{2}}{n\varphi_{x}(\underline{h}\lambda)\xi_{1}^{2}},
\label{variance}
\end{align}
where $\lambda$ and $\widehat{m}_2(\cdot)$ are as in (\ref{q3}) and (\ref%
{fraction}), respectively, and $a_n\simeq b_n$ means $a_n=b_n+o(1)$. With
these results we are now ready to construct the asymptotic normality of our
estimator. 

\subsubsection{Limiting distribution under independence}

We first consider the situation in which there is a set of independent
exogenous regressors in the static regression context; that is, when
marginal regressors $X_s$ are i.i.d., satisfying Assumption D1. In this
case, the asymptotic normality can be established for a wide range of
different choices of distributions.

Recall that under Assumption B2, the distribution function $F$ of $X^2$ is
regularly varying with the index of variation $\rho<0$. Then by the
characterisation theorem of Karamata (1933) (see for example Feller (1971)),
there always exists a slowly varying function $\ell(x)$ that satisfies 
\begin{equation}
F(1/x)=x^{\rho}\ell(x).  \label{karamata}
\end{equation}
Now fix some $p$, the increment constant for bandwidth in Assumption B3, and
denote by $\mathcal{L}(t)$ the Laplace transform of $X^2$. Then we define
the following constants 
\begin{align*}
&\quad\quad\qquad\qquad\qquad\qquad\qquad C_{\ell}=\lim_{h\rightarrow0}\Big[%
\ell^{-1/2}\Big(h^{-\frac{4p}{2p-1}}\Big)\Big]  \notag \\
&C^{*}=\frac{(2\pi)^{(1+2 p\rho)}(2p-1)}{\Gamma^{-1}(1-\rho)\cdot (2p)^{%
\frac{2p(\rho+2)-1}{2p-1}}}\cdot \zeta^{\frac{2p(1+\rho)}{2p-1}},\quad\quad
C^{**}=(2p-1)\cdot\left(\frac{K}{2p}\right)^{2p/(2p-1)}
\end{align*}
and 
\begin{equation*}
\zeta=-\int_0^{\infty}\frac{u^{-1/2p}\mathcal{L}^{\prime }(u)}{\mathcal{L}(u)%
}du,\quad\text{and}\quad\mathcal{V}_1=\frac{C^{*}C_{\ell}\xi_2\sigma^2(x)}{%
p^{*}(0)\xi_1^2\lambda^{\frac{1+2\rho p}{2p-1}}},
\end{equation*}
where $\Gamma(\cdot)$ is the Gamma function, $\xi_{1}$ and $\xi_{2}$ are the
constants specified in (\ref{q10}), $\lambda$ is the upper bound of the
support of the kernel, $p^{*}(\cdot)$ is the Radon-Nikodym derivative in B1.
As pointed out by Dunker et al. (1999), the constant $\zeta$ may not exist
in some cases. 

To aid the exposition, we compute the constants for some common, regularly
varying distributions in Table 1 below.

\begin{table}[h]
\centering
\begin{tabular}{c|c|c|c}
$X_j^2\sim F$ i.i.d. & $\rho$ & $\lim_{x\rightarrow\infty}\ell(x)=C_{%
\ell}^{-2}$ & $\zeta$ \\ \hline
Uniform(1,b) & $-1$ & $1$ & n/a \\ 
Gamma$(\alpha,\beta)$ & $-\alpha$ & $\beta^{\alpha}\alpha^{-1}\Gamma(%
\alpha)^{-1}$ & $\frac{\alpha\pi\beta^{-1/2p}}{\sin(\pi/2p)}$ \\ 
exp$(\eta)$ & $-1$ & $\eta$ & $\frac{\pi\eta^{-1/2p}}{\sin(\pi/2p)}$ \\ 
Weibull$(\alpha,\beta)$ & $-\alpha$ & $\beta$ & n/a \\ 
Pareto$(\theta,\mu)$ & $-1$ & $\mu/\theta$ & n/a \\ 
$\chi_1^2$ & $-1/2$ & $(2/\pi)^{1/2}$ & $\frac{\pi2^{(1-2p)/2p}}{\sin(\pi/2p)%
}$ \\ \hline
\end{tabular}%
\caption{Examples of some key constants for common distributions}
\end{table}

\noindent The main result of this subsection now follows. The theorem
derives the limiting distribution of the infinite-dimensional
Nadaraya-Watson type estimator for every i.i.d. regressor satisfying B2, and
a set of sufficient regularity conditions under which the expression is
valid. Notice that the specific choice of the regressor does not affect the
rate of convergence. \newline

\noindent\textsc{Theorem 2}.\emph{\ Suppose A2-A8 and B1-B4 hold. Let the
marginal regressors $X_s$ satisfy Assumption D1. Then the estimator (\ref%
{q12}) with respect to the sample observations }$\{Y_{t},X_{t}^{\intercal}%
\}_{t=1}^{n}$\emph{\ satisfying S1 is asymptotically normal with the
following limiting distribution: } 
\begin{align}
\sqrt{nh^{\frac{1+2\rho p}{2p-1}}\exp\Big\{-C^{**}(\lambda h)^{-\frac{2}{2p-1%
}}\Big\}\mathcal{V}_1^{-1}}\bigg[&\widehat{m}(x)-m(x) -\mathcal{B}_n \bigg] %
\Longrightarrow N\left(0, 1\right),  \label{q15}
\end{align}
\emph{where $\mathcal{B}_n=O(h^{\beta})$ is is the bias component as in (\ref%
{bias}).}



\subsubsection{Limiting distribution under gaussianity and dependence}

The strict independence condition between the regressors assumed in the
previous section can be relaxed to allow mild dependence specified in
Assumption D2. In doing so, we shall make use of the arguments used in Hong,
Lifshits and Nazarov (2015, Theorem 1), where the asymptotics of the small
deviation probability is investigated under dependence. This setting not
only grants sufficient flexibility in the static regression case, but
moreover allows distributional result for the autoregressive context where
the regressors constitute time lags of the response $X_s=Y_{t-s}$, with
dependence structure as stipulated in Assumption S2 (and thereby satisfying
D2). The cost required for this modification is Gaussianity assumption on
the regressor.

With reference to Table 1, we can easily compute the constants $C^{*}$ and $%
C^{**}$ for the Gaussian case, denoted $C^{*}_G$ and $C^{**}_G$
respectively, as follows: 
\begin{align*}
C^{*}_G=\frac{(2\pi)^{(1-p)}(2p-1)}{2\cdot (2p)^{\frac{3p-1}{2p-1}}}\cdot %
\left[\frac{\pi2^{(1-2p)/2p}}{\sin(\pi/2p)}\right]^{\frac{-p}{2p-1}%
},\quad\quad C^{**}_G=\frac{2p-1}{2}\left( \frac{\pi }{2p\sin{\frac{\pi}{2p}}%
}\right)^{\frac{2p}{2p-1}}.
\end{align*}

With other constants defined as before, we now have the following asymptotic
normality for the case of dependent regressors. We reiterate that the result
covers both the regression and autoregression context (S1 and S2), and is
invariant as long as the cross-sectional dependence structure satisfies
Assumption D2. 

For the square summable sequence $a_{j}$ in (\ref{crsec}) define 
\begin{equation*}
C_{\mathcal{A}}=\left[ \frac{1}{2\pi }\int_{0}^{2\pi }\bigg|%
\sum_{j=0}^{\infty }a_{j}\exp (\text{i}jx)\bigg|^{1/p}~dx\right] ^{p}\quad 
\text{and}\quad \mathcal{V}_{2}=\frac{C_{G}^{\ast }C_{\ell }\xi _{2}\xi
_{1}^{-2}\sigma ^{2}(x)}{e^{-\frac{1}{2}\Vert \Gamma ^{-1/2}z\Vert
_{2}^{2}}(C_{\mathcal{A}}\lambda )^{\frac{1-p}{2p-1}}}.
\end{equation*}%
where $\sigma ^{2}(\cdot )$ is the conditional variance defined in
Assumption B4 and $z=(z_{j})=(j^{-p}x_{j})$. Then we have the following
result:\\

\noindent \textsc{Theorem 3}. \emph{Suppose A2-A8 and B1-B4 hold. Let the
regressor }$X=(X_{1},X_{2},...)^{^{\intercal }}$\emph{\ is jointly normally
distributed with zero mean and the covariance operator }$\Gamma $\emph{, and
satisfies D2. Then, the estimator (\ref{q12}) with respect to sample
observations $\{Y_{t},X_{t}^{\intercal }\}_{t=1}^{n}$ satisfying either S1
or S2 is asymptotically normal with the following limiting distribution:} 
\begin{equation}
\sqrt{nh^{\frac{1-p}{2p-1}}\exp \Big\{-C_{G}^{\ast \ast }(C_{\mathcal{A}%
}\lambda h)^{-\frac{2}{2p-1}}\Big\}\mathcal{V}_{2}^{-1}}\bigg[\widehat{m}%
(x)-m(x)-\mathcal{B}_{n}\bigg]\Longrightarrow N\left( 0,1\right) ,
\label{q17}
\end{equation}%
%
\emph{where }$\mathcal{B}_{n}$\emph{\ is again the bias component in (\ref%
{bias}).}\newline
\newline
\indent\textsc{Remark.} The additional term $C_{\mathcal{A}}$ is a function
of the sequence $a_{j}$ that reflects the dependence structure allowed
between the regressors. This can be thought of as the \textquotedblleft
penalty term\textquotedblright\ for allowing 
dependence. The exponential term in the denominator of the asymptotic
variance arises from the asymptotic equivalence relationship between the
shifted and non-shifted small deviation for $\ell _{2}$-valued Gaussian
variables, see (\ref{zolo2}) in the appendix.


\subsection{Optimal Bandwidth}

We now briefly discuss the issue of bandwidth choice. As in the finite
dimensional framework, the results above confirm the existence of the
trade-off relationship between the order of bias and variance. As the
bandwidth grows, the variance decreases and the bound on the bias increases,
vice versa. 
Therefore we may search for the optimal bandwidth $h_{opt}$ that balances
the order of those two quantities by solving their equivalence relation.

For example, as for the case of Gaussian regressor we have 
\begin{equation}
h^{\beta}\sim\sqrt{\frac{\exp\big(Ch^{-2/(2p-1)}\big)}{nh^{\frac{1-p}{2p-1}}}%
}
\end{equation}
so that 
\begin{align}
\left[ 2\beta+\frac{1-p}{2p-1}\right] \cdot\log h-Ch^{-\frac {2}{2p-1}}
&\sim -\log n.  \notag
\end{align}
Taking $h\sim(\log n)^{a}$ for some strictly negative $a$ balances the
leading terms on both sides: 
\begin{equation}
\left[ 2\beta+\frac{1-p}{2p-1}\right] \cdot a\cdot\log\log n-(\log n)^{-%
\frac{2}{2p-1}\cdot a}\sim -\log n.  \label{qz1}
\end{equation}
The explicit order $a$ that solves (\ref{qz1}) can be computed in terms of $%
n $, $\beta$ and $p$. Writing $j:=\left[ 2\beta+(1-p)/(2p-1)\right] $ and $%
k:=2/(2p-1)$ for the sake of simplicity, we can show after some computation
that 
\begin{equation}
a_{opt}=\frac{j\cdot \mathcal{W}\left(\frac{k}{j}\cdot n^{k/j}\right) -k\log
n}{jk\cdot\log\log n},  \label{q26}
\end{equation}
where $\mathcal{W}(y)$ is the lambert W function, which returns the solution 
$x$ of $y=x\cdot e^{x}$. From this result the optimal bandwidth $h_{opt}\sim
(\log n)^{a_{opt}}$ follows.\newline

\textsc{Remark.} We can search for the optimal bandwidth for the cases of
non-Gaussian regressors by following exactly the same manner as above, so
tedious details are omitted here. As regards the solution in (\ref{q26}),
since the mapping $x\mapsto x\cdot e^{x}$ is not injective, the solution may
be multi-valued on the negative domain, i.e. $y<0$. This does not happen in (%
\ref{q26}) provided $\beta\geq1/4$ (however big $p$ is), because $%
(1-p)/(2p-1)$ is bounded away from $-1/2$. .\newline

Since the $\log$ terms dominate the double logarithm in (\ref{qz1}) as the
number of sample $n$ increases, it can be readily expected that the optimal
value of $a$ in (\ref{q26}) converges to a limit in such a way that the
leading orders are balanced. Below we introduce without formal justification
a result that gives the lower bound (infimum) of the optimal bandwidth.
Obviously, the result holds for any choice of the distribution of the
regressor, since the order of the leading terms $-2/(2p-1)$ remains
invariant as was shown in (\ref{q15}) and (\ref{q17}). \newline
%

\noindent\textsc{Corollary 2.}\emph{\ For any fixed choice of }$p>1$\emph{\
and the distribution $F$ of $X^2$ satisfying B2, the order of the optimal
bandwidth $a_{opt}$ satisfies} 
\begin{equation}
a_{opt}\downarrow\left(-\frac{2p-1}{2}\right)\quad\text{as}\quad
n\rightarrow\infty,
\end{equation}
\emph{which suggests that the lower bound of the optimal bandwidth is given
by\footnote{%
The notation $f\preceq g$ means there exists some positive constant $c$ such
that $\lim_{n\rightarrow\infty}f(n)/g(n)\leq c$.}} 
\begin{equation}
(\log n)^{-\frac{2p-1}{2}}\preceq h_{opt}\sim (\log n)^{a_{opt}}.
\end{equation}

\subsection{Uniform consistency}

Uniform consistency of the Nadaraya-Watson estimator was first studied by
Nadaraya (1964, 1970) and subsequently by numerous others. To name a few
earlier literature, Devroye (1978) moderated the regularity conditions
required in the previous papers, and Robinson (1983) proved consistency for
dependent sample data. In functional statistics literature, only uniform
consistency with respect to i.i.d. sample has been established so far, see
Ferraty et al. (2010). Following their approach, we introduce the notion of
Kolmogorov's entropy which accounts for the complexity of infinite
dimensional spaces.\newline

\noindent\textsc{Definition 5}\emph{. Given some }$\eta>0$\emph{, let }$%
L(S,\eta)$ \emph{be the smallest number of open balls in }$E$\emph{\ of
radius }$\eta$\emph{\ needed to cover the set }$S\subset E$\emph{. Then
Kolmogorov's }$\eta$\emph{-entropy is defined as }$\log L(S,\eta)$.\newline

From the definition it can be readily expected that Kolmogorov's entropy
depends heavily on the nature/structure of the space we work on, and
therefore is closely related to the rate of convergence of the estimator.

It is well known that the regression function cannot be estimated uniformly
over the entire space because the magnitude of the regression function $%
m(\cdot)$ becomes non-predictable as $\|x\|\rightarrow\infty$, see Bosq
(1996). In our framework, even greater restrictions apply; since we are
working on infinite sequence spaces, none of their subsets can be covered by
a finite number of balls, and $L(S,\eta)$ then becomes infinite. Therefore,
we propose to adopt an approximation argument and consider uniform
consistency over a set whose effective dimension is increasing in sample
size $n$. In particular, we define the set 
\begin{equation}
S_{\tau}:=\big\{u|(u_{i})_{i\in\mathbb{Z}^{+}}, u_{j}=0\text{ for all }%
j>\tau, \|u\|_{\infty}\leq\lambda\big\}  \label{q18}
\end{equation}
where $\tau=\tau_n$ is some increasing sequence and fixed real number $%
\lambda$, and consider uniform consistency over this compact set.
Kolmogorov's entropy of the set $S_{\tau}$ is given as follows:\newline

\noindent \textsc{Lemma 2}\emph{. Kolmogorov's }$\eta $\emph{-entropy of} $%
S_{\tau }$\emph{\ defined in (\ref{q18}) with }$\tau =\tau _{n}(\rightarrow
\infty )$\emph{\ and }$\lambda >0$\emph{\ is} 
\begin{equation}
\log L\big(S_{\tau },\eta \big)=\log \Bigg[\Bigg(\frac{2\lambda \sqrt{\tau }%
}{\eta }+1\Bigg)^{\tau }~\Bigg].  \label{q19} \\
\end{equation}%
\newline
\indent\textsc{Remark.} We note that (\ref{q19}) is indeed in line with
common intuition; as the dimension $\tau $ increases, the number of balls
with some fixed radius required to cover the set goes off to infinity. The
proof of this result can be done by exploiting the splitting technique and
then by attempting to cover the polyhedron of increasing dimension. See
appendix for details. From this result it readily follows that for fixed $%
\delta $ and $\lambda $, Kolmogorov's entropy $\log L(S_{\tau },\eta )$ is
of order $O(\tau +\tau \cdot \log \tau )=O(\tau \log \tau )$. As we shall
discuss below, this has some implications with regard to the choice of the
sequence $\tau _{n}$.

We now introduce some further assumptions needed for uniform consistency:\\


\noindent\textsc{Assumptions C}

\begin{enumerate}
\item[C1.] \emph{The Kolmogorov's }$\eta$\emph{-entropy }$\log L(S_{\tau
},\eta)$\emph{\ satisfies} 
\begin{equation}
\frac{(\log n)^{3+\epsilon}}{n\varphi_x(\underline{h})}<\log L(S_{\tau},\eta)<\frac{%
n\varphi_x(\underline{h})}{\log n}\quad\text{for some}~~\epsilon\in(0,1)  \label{q20}
\end{equation}
\emph{Furthermore,} $0<\varphi_x(\underline{h})\leq Ch<\infty$\emph{\ and } 
$(\log n)^2/(n\varphi_x(\underline{h}))\longrightarrow0$\emph{\ as }$n\rightarrow\infty$.

\item[C2.] \emph{The kernel function $K$ is Lipschitz continuous on $%
[0,\lambda]$.}
\end{enumerate}
\textrm{}\\
\textsc{Remark}. The first part of assumption C1 specifies the rate at which
Kolmogorov's entropy should behave in sample size $n$ (hence in dimension $%
\tau _{n}$) in order for the theory to hold. 
From the upper and lower bound it readily follows that $n\varphi (h)$ must
be of order strictly bigger than $(\log n)^{2+\epsilon /2}$. This assumption
is general and is nothing restrictive; for example, taking $h\sim (\log
n)^{-(2p-1)/2}$ in view of Corollary 2 and (27), it follows that $n\varphi
(h)\sim (\log n)^{(2p-1)\beta }$. In this case, assumption (\ref{q20}) is
valid as long as $p$ is moderately large enough relative to $\beta \leq 1$.
The second part of C1 is straightforward, and the last part only
slightly strengthens the bandwidth condition in A2. \newline

We now introduce the main result of this section. Note that with a slight
abuse of notation $X$ is hereafter taken to mean the regressor, but with
zeros after its $\tau^{th}(=\tau_{n}\rightarrow\infty~\text{as}%
~n\rightarrow\infty)$ entry; that is, 
\begin{equation*}
X=(X_{1},X_{2},...,X_{\tau},0,0,...)^{^{\intercal}}
\end{equation*}
(so that the original $X$ is recovered as $n\rightarrow\infty$).
Also, the regression operator and the estimator with respect to this
truncated regressor are denoted by $m_{\tau}(\cdot)$ and $\widehat{m}%
_{\tau}(\cdot)$, respectively. 

A stonger condition is needed on the rate of mixing; from hereafter, by S1' and S2' we mean assumptions S1 and S2 but with the arithmetric mixing rate condition strengthened to exponential mixing (cf. Definition 1) with some exponent $\varsigma>0$.\\

\noindent\textsc{Theorem 4.}\emph{\ Suppose that Assumptions A2-A8, B1-B3
and C1-C2 hold. Let the marginal regressors $X_s$ satisfy D1, and let $\tau=\tau_n\sim (\log n)$. Then the estimator }$\widehat {m}_{\tau}(\cdot)$\emph{\ with respect to sample
observations }$\{Y_{t},X_{t}\}_{t=1}^{n}$\emph{\ satisfying S1' is almost
sure uniformly consistent for }$m(x)=m(x_{1},...)$\emph{\ over }$S_{\tau}$%
\emph{:} 
\begin{equation}
\sup_{x\in S_{\tau}}\Big|\widehat{m}_{\tau}(x)-m_{\tau}(x)\Big|=O_{a.s.} %
\Bigg(h^{\beta}+ \sqrt{\frac{(\log n)^{2} \exp\Big\{\big(h^2+\tau^{1-2p}\big)%
^{^{-1/(2p-1)}}\Big\}}{n\big(h^2+\tau^{1-2p}\big)^{\frac{1+2p\rho}{2(2p-1)}}}%
}~\Bigg),  \label{f22}
\end{equation}
\emph{If alternatively $X_s$ is Gaussian and satisfies D2, then the same
conclusion holds with respect to sample observations satisfying either S1' or
S2':}\newline
\begin{equation}
\sup_{x\in S_{\tau}}\Big|\widehat{m}_{\tau}(x)-m_{\tau}(x)\Big|=O_{a.s.} %
\Bigg(h^{\beta}+ \sqrt{\frac{(\log n)^{2} \exp\Big\{\big(h^2+\tau^{1-2p}\big)%
^{^{-1/(2p-1)}}\Big\}}{n\big(h^2+\tau^{1-2p}\big)^{\frac{1-p}{2(2p-1)}}}}~%
\Bigg).  \label{q22} \\
\end{equation}

\textsc{Remark.} The $\tau $ terms in the convergence rate reflects the
penalty for truncation. Since $\tau \sim (\log n)$ the order of the term $%
h^{2}$ dominates that of $\tau ^{1-2p}$ as long as $h\succeq (\log
n)^{-(2p-1)/2}$. This can be done by choosing the bias-variance optimal
bandwidth; following the same arguments in the pointwise case, choosing $%
h\sim (\log n)^{a}$ and solving for $n$ we end up with 
\begin{equation*}
a_{opt}=\frac{j\cdot \mathcal{W}\left[ \frac{k}{j}\exp (-\frac{k}{j}2\log
\log n-k\log n)\right] +2k\log \log n-k\log n}{jk\log \log n}.
\end{equation*}%
And because the order of the leading terms is $(\log n)^{-(2p-1)/2}$ as in
the pointwise case, it is straightforward to see that the lower bound of the
optimal bandwidth in Corollary 2 still continues to hold; that is, $%
h_{opt}\succeq (\log n)^{-(2p-1)/2}$. This is again invariant to the choice
of distribution $F$ of the regressor. 

We end up with the result that follows. Note that $\beta$ is the measure of smoothness defined in assumption A8.\newline

\noindent\textsc{Corollary 3.} \emph{Suppose conditions assumed in Theorem 4 hold. Then, upon choosing }$h\sim(\log n)^{a_{opt}}$\emph{, we have }
\begin{equation}
\sup_{x\in S_{\tau}}\Big|\widehat{m}_{\tau}(x)-m_{\tau}(x)\Big|=O_{a.s.}\Big(%
\lbrack\log n]^{\beta\cdot a_{opt}}\Big).  \label{q23}
\end{equation}

\section{Appendix: Proofs of the main results}

Throughout, $C$ (or $C^{\prime }$, $C^{\prime \prime }$) is taken to mean
some generic constant that may take different values in different places
unless defined specifically otherwise.

\subsection{Proof of Theorem 1}

Recalling the decomposition (\ref{decomp1}): 
\begin{align*}
\widehat{m}(x)-m(x) = \frac{E\widehat{m}_{2}(x)-m(x)}{\widehat{m}_{1}(x)}+ 
\frac{\widehat{m}_{2}(x)-E\widehat {m}_{2}(x)}{\widehat{m}_{1}(x)}-\frac{%
m(x)[\widehat{m}_{1}(x)-1]}{\widehat{m}_{1}(x)},
\end{align*}
we shall show $E\widehat{m}_{2}(x)-m(x)\rightarrow 0$ and $\widehat{m}%
_{2}(x)-E\widehat{m}_{2}(x)\rightarrow^{p}0$, since $\widehat{m}%
_1(x)\rightarrow^p 1$ will then follow from the latter and complete the
proof.

As regards the first `bias component', denoting by $\mathcal{E}(x,\lambda%
\underline{h})$ the infinite dimensional hyperellipsoid centred at $x\in%
\mathbb{R}^{\mathbb{N}}$ with semi-axes $h_{j}$ in each direction we have 
\begin{align}
E\widehat{m}_{2}(x)-m(x) & =E\Bigg(\frac{1}{nEK_{1}}%
\sum_{t=1}^{n}K_{t}Y_{t}-m(x)\Bigg)  \notag \\
& =\frac{1}{EK_{1}}EK_{1}Y_{1}-\frac{EK_{1}}{EK_{1}}m(x)=\frac{1}{EK_{1}}E%
\bigg[E\bigg[\big(Y_{1}-m(x)\big)K_{1}\Big|X\bigg]\bigg]  \notag \\
& =\frac{1}{EK_{1}}E\Big[\Big[m(X)-m(x)\Big]K_{1}\Big]\leq\sup_{u\in 
\mathcal{E}(x,\lambda\underline{h})}\big|m(u)-m(x)\big|\longrightarrow 0
\label{q29}
\end{align}
as $n\rightarrow\infty$, where $K_t$ is the shorthand notation for $%
K(\|H^{-1}(x-X_{t})\|)$ as introduced in the main text above. 
The second equality is justified by stationarity that is preserved under
measurable transformation, and the last inequality is due to compact support
of the kernel and continuity of the regression operator at $x$ (Assumption
A1). 

The next step concerns with the `variance component', where the mean-squared
convergence of $\widehat{m}_{2}-E\widehat{m}_{2}$ to zero will be shown. We
write 
\begin{equation}
\widehat{m}_{2}-E\widehat{m}_{2}=\frac{1}{n}\sum_{t=1}^{n}\frac{1}{EK_{1}}%
\bigg\{K_{t}Y_{t}-E(K_{t}Y_{t})\bigg\}=:\frac{1}{n}\sum_{t=1}^{n}Q_{t}.
\label{mmp}
\end{equation}%
The arguments to follow depend on the temporal dependence structure of $Q_{t}$. 
In the static regression case, $Q_{t}$ is a
measurable function of $Y_{t},X_{1t},X_{2t},\ldots $, and hence inherits their
joint dependence structure. That is, $Q_{t}$ is also arithmetically $\alpha 
$-mixing with the rate specified in S1. In the autoregression case, we assume near
epoch dependence of $K_{t}$ (and hence $Q_t$; upon assuming a slightly stronger condition on the stability coefficient) on $Y_t$ as specified in Assumption S2, due to its dependence upon infinite past of $Y_{t}$. We shall hence deal with two cases separately. 
\newline

\noindent\textsc{Case 1: Static Regression.} It is clearly sufficient to
prove $\text{Var}(\widehat{m}_{2}-E\widehat{m}_{2})\rightarrow0$ to show
mean squared convergence. Since $Q_{t}$ is (temporally) stationary we have 
\begin{align}
\text{Var}(\widehat{m}_{2}-E\widehat{m}_{2}) & =\frac{1}{n^{2}}\sum_{t=1}^{n}%
\text{Var}(Q_{t})+\frac{2}{n^{2}}\mathop{\sum\sum}_{1\leq i<j\leq n} \text{%
Cov}\big(Q_{i},Q_{j}\big)  \label{q31} \\
& =\frac{1}{n}\text{Var}(Q_{1})+\frac{2}{n^{2}}\mathop{\sum\sum}%
_{1\leq|i-j|<n} \text{Cov}\big(Q_{i},Q_{j}\big)  \notag \\
& =\frac{1}{n}\text{Var}(Q_{1})+\frac{2}{n^{2}}\sum_{s=1}^{n-1}(n-s)\cdot 
\text{Cov}\big(Q_{1},Q_{s+1}\big)  \notag \\
& =:A_{1}+A_{2}.  \label{q32}
\end{align}

\noindent Now, by (\ref{q7}), (\ref{q8}) and Assumption A it follows that 
\begin{align}
~A_{1} & =\frac{1}{nE^{2}K_{1}}\text{Var}\bigg(K_{1}Y_{1}-EY_{1}K_1\bigg)=%
\frac{\text{Var}\left(K_{1}Y_{1}\right)}{nE^{2}K_{1}}\leq\frac{EK_1^2Y_1^2}{%
nE^{2}K_{1}}  \notag \\
& =\frac{EK_1^2Y_1^21\{|Y|\leq \upsilon\}}{nE^{2}K_{1}}+\frac{%
EK_1^2Y_1^21\{|Y|> \upsilon\}}{nE^{2}K_{1}}\leq \frac{C\upsilon^2}{%
n\varphi_x(\lambda \underline{h})}+\frac{CE|Y|^{2+\delta}\upsilon^{-\delta}}{%
n\varphi_x(\lambda \underline{h})},  \label{q33}
\end{align}
where $\upsilon=\upsilon_n$ is some increasing truncation sequence. Then (%
\ref{q33}) tends to zero if the sequence is chosen to increase sufficiently
slowly in such a way that $\upsilon=o(\sqrt{n\varphi})$. 


We now move on to the second term $A_{2}$ and investigate the covariance
term. 
Since measurable transformations of mixing variables preserve the mixing
property, 
using Davydov's inequality, see Davydov (1968, Lemma 2.1) or Bosq (1996, Corollary 1.1), and boundedness of the kernel functions (\ref{q3}%
) we have 
\begin{align}
\left|\text{Cov}\big(Q_{1},Q_{s+1}\big)\right|&=\Bigg|\text{Cov}\Bigg(%
Y_{1}\frac{K_{1}}{EK_{1}},Y_{s+1}\frac{K_{s+1}}{EK_{1}}\Bigg)\Bigg|\notag\\
&\leq \frac{12C\cdot\{E|Y_t|^{2+\delta}\}^{\frac{2}{2+\delta}}s^{-k\delta/(2+%
\delta)}}{\varphi _{x}(\underline{h}\lambda)^{2}}.  \label{q34}
\end{align}
In the meantime, 
\begin{align}
\big|\text{Cov}\big(Q_{1},Q_{s+1}\big)\big| &=\Bigg|\text{Cov}\Bigg(Y_{1}%
\frac{K_{1}}{EK_{1}},Y_{s+1}\frac{K_{s+1}}{EK_{1}}\Bigg)\Bigg|  \notag \\
& \leq\Bigg|E\Bigg(Y_{1}\frac{K_{1}}{EK_{1}}Y_{s+1}\frac{K_{s+1}}{EK_{1}}%
\Bigg)\Bigg|+\Bigg|E\Bigg(Y_{1}\frac{K_{1}}{EK_{1}}\Bigg)E\Bigg(Y_{s+1}\frac{%
K_{s+1}}{EK_{1}}\Bigg)\Bigg|  \notag \\
& \leq\frac{C}{\varphi_x(\underline{h}\lambda)^2}\left|E\left(K_{1}K_{s+1}%
\right)\right|+\frac{C}{E^2K_1}\left|E\left(K_1\right)E\left(K_{s+1}\right)%
\right|  \notag \\
& \leq\frac{C}{\varphi_x(\underline{h}\lambda)^2}\cdot\psi_x(\lambda%
\underline{h};1,s+1)+C^{\prime }\leq C^{\prime \prime }  \label{q35}
\end{align}
by stationarity, law of iterated expectation, boundedness of regression
function, and Assumption A6, A5 (along with the upper bound $\psi(\lambda%
\underline{h};1,s+1)$ of $EK_1K_{s+1}$ obtained as a direct consequence of
A5 following similar arguments used for Lemma 1).

With reference to (\ref{q34}) and (\ref{q35}), we take some increasing
sequence $r_{n}\rightarrow\infty$ such that $r_n=o(n)$, and write 
\begin{align}
\sum_{s=1}^{n-1}\big|\text{Cov}\big(Q_{1},Q_{s+1}\big)\big| &
=\sum_{s=1}^{r_{n}-1}\big|\text{Cov}\big(Q_{1},Q_{s+1}\big)\big|%
+\sum_{s=r_{n}}^{n-1}\big|\text{Cov}\big(Q_{1},Q_{s+1}\big)\big|  \notag
\\
& \leq C^{\prime \prime }\big(r_{n}-1)+\sum_{s=r_{n}}^{n-1}\frac{%
Cs^{-k\delta/(2+\delta)}}{\varphi _{x}(\underline{h}\lambda)^{2}}= O\bigg(%
r_{n}+\frac{r_{n}^{-k\delta/(2+\delta)+1}}{\varphi _{x}(\underline{h}%
\lambda)^{2}}\bigg),  \label{q36}
\end{align}
which is $O\big(\varphi_{x}(\underline{h}\lambda)^{-2(2+\delta)/(k\delta)}%
\big)$ upon choosing $r_{n}\sim\varphi_{x}(\underline{h}\lambda)^{-2(2+%
\delta)/(k\delta)}$.

\noindent Consequently, since $k\geq 2(2+\delta)/\delta$ it follows that 
\begin{align}
A_{2}:=& ~\frac{2}{n^{2}}\sum_{s=1}^{n-1}(n-s)\cdot \text{Cov}\big(%
Q_{1},Q_{s+1}\big)=\frac{2}{n}\sum_{s=1}^{n-1}\bigg(1-\frac{s}{n}\bigg)%
\cdot \text{Cov}\big(Q_{1},Q_{s+1}\big)  \notag \\
=&~ O\bigg(\frac{1}{n[\varphi_{x}(\underline{h}\lambda)]^{2(2+\delta)/(k%
\delta)}}+\frac{1}{n^{2}[\varphi_{x}(\underline{h}\lambda)]^{2(2+\delta)/(k%
\delta)}}\bigg)  \notag \\
=&~O\bigg(\frac {1}{n[\varphi_{x}(\underline{h}\lambda)]^{2(2+\delta)/(k%
\delta)}}\bigg)\longrightarrow0\quad\text{as}\quad n\rightarrow\infty
\label{q37}
\end{align}
and the desired result is obtained. \\

\noindent\textsc{Case 2: Autoregression.} We return back to (\ref{mmp}): 
\begin{equation}
\widehat{m}_{2}-E\widehat{m}_{2}=\frac{1}{n}\sum_{t=1}^{n}\frac{1}{EK_{1}}%
\bigg\{K_{t}Y_{t}-E(K_{t}Y_{t})\bigg\}=:\frac{1}{n}\sum_{t=1}^{n}Q_{t}.
\label{z1}
\end{equation}
In this framework $K_t=K(\|H^{-1}(x-X_t)\|)$ is a function of $%
(Y_{t-1},Y_{t-2},\ldots)$; despite loosing the mixing property, it inherits
stationarity of the mixing process $\{Y_t\}$. 
We write $K_{t,(m)}=\Psi_m(Y_t,Y_{t-1},Y_{t-2},\ldots,Y_{t-m+1})=E(K_t|Y_t,\ldots,Y_{t-m+1})$, where $m$ is as in assumption S2. Clearly, $K_{t,(m)}$ preserves the mixing dependence structure of $Y_t$ by elementary arguments, see e.g. Lu (2001, Remark 1(b)). Now write 
\begin{align}
\widehat{m}_{2}-E\widehat{m}_{2} &=\frac{1}{n}\sum_{t=1}^{n}\frac{1}{EK_{1}}%
\bigg[K_{t,(m)}Y_t-E\big(K_{t,(m)}Y_{t}\big)\bigg]+\frac{1}{n}\sum_{t=1}^{n}%
\frac{1}{EK_{1}}\bigg[K_{t}Y_{t}-K_{t,(m)}Y_t\bigg]  \notag \\
&\qquad\quad+~\frac{1}{n}\sum_{t=1}^{n}\frac{1}{EK_{1}}\bigg[E\big(%
K_{t,(m)}Y_{t}\big)-E(K_{t}Y_{t})\bigg]=R_1+R_2+R_3,  \label{autor}
\end{align}
and first consider the last term $R_3$.

Fix some increasing sequence $q=q_n\rightarrow\infty$, and write $%
Y_{t,L}:=Y_t1\{|Y_t|\leq q\}$ and $Y_{t,U}=Y_t1\{|Y_t|> q\}$. Then 
\begin{align}
EY_tK_{t,(m)}&=EY_tK\big(\|H^{-1}(x-X_t)\|\big)-EY_{t,U}K\big(%
\|H^{-1}(x-X_t)\|\big)  \notag \\
&\qquad+EY_{t,L}K_{t,(m)}-EY_{t,L}K\big(\|H^{-1}(x-X_t)\|\big)  \notag \\
&\qquad+EY_{t,U}K_{t,(m)}  \notag \\
&=D_1+D_2+D_3.  \label{pd1}
\end{align}
The second part of $D_1$ is given by 
\begin{align}
EY_{t,U}K\big(&\|H^{-1}(x-X_{t})\|\big)\leq E|Y_t|1_{\{|Y_t|>q\}}K\big(%
\|H^{-1}(x-X_t)\|\big)  \notag \\
&\leq q^{-(\delta+1)}E|Y_t|^{2+\delta}1_{\{|Y_t|>q\}}K_t  \notag \\
&\leq Cq^{-(\delta+1)}E|Y_t|^{2+\delta}1_{\{|Y_t|>q\}}=o(q^{-(\delta+1)})
\label{pd2}
\end{align}
because $1_{\{|Y_t|>q\}}=o(1)$ as $n\rightarrow\infty$. Following similar
arguments on $D_3$ we have $D_1+D_3=EY_tK_t+o(q^{-(\delta+1)})$. So we are now left with the middle term $D_2$: 
\begin{align}
D_2&\leq E\left|Y_{t,L}\right|\left|K_t-K_{t,(m)}\right|
=O\left(q\sqrt{v_2(m_n)}\right)  \label{pd3}
\end{align}
by H\"{o}lder's inequality. Therefore in view of (\ref{pd1}),(\ref{pd2}) and (\ref{pd3}) we have 
\begin{equation}
R_3=\frac{1}{nEK_1}\sum_{t=1}^{n}\bigg[EK_{t,(m)}Y_t-E(K_tY_t)\bigg]=o\left(%
\frac{q^{-(\delta+1)}}{\varphi_x(\lambda\underline{h})}\right)+O\left(\frac{q%
\sqrt{v_2(m_n)}}{{\varphi_x(\lambda\underline{h})}}\right),
\end{equation}
and upon choosing $q=(\varphi_x(\underline{h}\lambda)/n)^{-1/(2(\delta+1))}$
we have $o(\varphi^{-1}q^{-(\delta+1)})=o(\varphi^{-1}(%
\varphi/n)^{1/2})=o(n^{-1/2}\varphi^{-1/2})=o(1)$, and also 
\begin{align}
O\left(\frac{1}{\varphi_x(\underline{h}\lambda)}q\sqrt{v_2(m_n)}%
\right)&=O\left(\frac{1}{\varphi_x(\underline{h}\lambda)}\cdot\left(\frac{%
\varphi_x(\underline{h}\lambda)}{n}\right)^{-1/(2(\delta+1))}\sqrt{v_2(m_n)}%
\right)  \notag \\
&=O\left(\frac{\sqrt{v_2(m_n)}}{[\varphi_x(\underline{h}\lambda)]^{(2%
\delta+3)/(2\delta+2)}n^{-1/(2(\delta+1))}}\right)=o(1)  \label{autor1}
\end{align}

\noindent by Assumption S2. Hence $R_{3}=o(1)$ and similarly $R_{2}=o_{p}(1)$.

As for the first term that remains, we can rewrite $R_{1}$ as 
\begin{equation*}
\frac{1}{n}\sum_{t=1}^{n}\Bigg[\frac{K_{t,(m)}Y_{t}-E(K_{t,(m)}Y_{t})}{%
EK_{1,(m)}}\Bigg]+\frac{1}{n}\sum_{t=1}^{n}\Bigg[\frac{%
\{K_{t,(m)}Y_{t}-E(K_{t,(m)}Y_{t})\}\cdot \{EK_{1,(m)}-EK_{1}\}}{%
EK_{1,(m)}\cdot EK_{1}}\Bigg],
\end{equation*}%
and denote by each summation as $F_{1}$ and $F_{2}$, respectively.

On noting that $K_{t,(m)}$ is $\alpha $-mixing, we follow the arguments we
used for the static regression case. Write $%
(K_{t,(m)}Y_{t}-E(K_{t,(m)}Y_{t}))/EK_{1,(m)}=Q_{t,(m)}$; then since $Q_t$ is near epoch dependent on $Y_t$ thanks to (the extra factor $\varphi^{-1}$ in) condition (\ref{f2a}) stipulated on the rate of stability, by using the approximation scheme one can readily see that 
the first component $F_{1}\rightarrow ^{p}0$.

Furthermore, moving on to the next term we see that 
\begin{align}
F_2&\leq\left[\frac{1}{n}\sum_{t=1}^{n}\frac{%
\{K_{t,(m)}Y_t-E(K_{t,(m)}Y_{t})\}}{EK_{1,(m)} }\right]\cdot\frac{C\sqrt{%
v_2(m_n)}}{EK_1}=o_p\left(\frac{\sqrt{v_2(m_n)}}{\varphi_x(\underline{h}%
\lambda)}\right),  \label{p10}
\end{align}
completing the proof. \hfill\mbox{\ \rule{.1in}{.1in}} 

\subsection{Proof of Theorem 2 and 3}
With additional sumptions A7, A8 and B3 it follows that 
\begin{align}
\mathcal{B}_n&=E\widehat{m}_{2}(x)-m(x) =E\bigg(\frac{1}{nEK_{1}}\sum_{t=1}^{n}K_{t}%
Y_{t}-m(x)\bigg)\nonumber\\
& =\frac{1}{EK_{1}}EK_{1}Y_{1}-\frac{EK_{1}}{EK_{1}}m(x)=\frac{1}{EK_{1}%
}E\bigg[E\bigg[\big(Y_{1}-m(x)\big)K_{1}\Big|X\bigg]\bigg]\nonumber\\
& =\frac{1}{EK_{1}}E\Big[\Big[m(X)-m(x)\Big]K_{1}\Big]\leq\sup_{u\in
\mathcal{E}(x,\lambda\underline{h})}\big|m(u)-m(x)\big|\nonumber\\
&\leq \sup_{u\in
\mathcal{E}(x,\lambda\underline{h})}\sum_{j=1}^{\infty}c_{j}\big|u_j-x_j\big|^{\beta}=\sum_{j=1}^{\infty}c_j(\lambda h\phi_j)^{\beta}= h^\beta\bigg(\lambda^\beta\sum_{j=1}^{\infty}c_jj^{p\beta}\bigg)<\infty.\label{asybias}
\end{align}
\noindent Moving on, rewriting the decomposition (\ref{decomp1}) as
\begin{equation*}
\widehat{m}(x)-m(x)-B(x)=\frac{B(x)\cdot\big[1-\widehat{m}_{1}(x)%
\big]}{\widehat{m}_{1}(x)}+\frac{\widehat
{m}_{2}(x)-E\widehat{m}_{2}(x)-m(x)\big[\widehat{m}_{1}(x)-1\big]}{\widehat{m}_{1}(x)},
\end{equation*}
we see that it suffices to derive the limiting distribution of 
\begin{align}
\widehat
{m}_{2}(x&)-E\widehat{m}_{2}(x)-m(x)[\widehat{m}_{1}(x)-1]\notag\\
&=\frac{1}{n}\sum_{t=1}^{n}\frac{1}{EK_{1}}\Big[K_{t}Y_{t}-m(x)K_{t}%
-E(K_{t}Y_{t})+m(x)EK_{t}\Big]=:\frac{1}{n}\sum_{t=1}^{n}R_{nt},\label{asdf}
\end{align}
since $\widehat{m}_1(x)\rightarrow^p 1$ as a consequence of Theorem 1. Now, with Assumption A6, B3, B4, and the law of iterated expectations, the asymptotic variance of (\ref{asdf}) is given by
\begin{align}
\frac{\textrm{Var}[K_{t}(Y_{t}-m(x))]}{nE^{2}K_{1}}&=\frac
{1}{nE^{2}K_{1}}\bigg\{E\Big[K_{t}(Y_{t}-m(x))\Big]^{2}-E^{2}\big[K_{t}%
(Y_{t}-m(x))\Big]\bigg\}\nonumber\\
& =\frac{1}{nE^{2}K_{1}}\Bigg\{E\big[\sigma^{2}(X)K_{1}^{2}\big]+E\bigg(\Big[m(X)-m(x)\Big]^2K_{1}^{2}\bigg)\Bigg\}\nonumber\\
& =\frac{1}{nE^{2}K_{1}}\Bigg\{\sigma^{2}(x)EK_{1}^{2}+E\bigg(\Big[\sigma
^{2}(X)-\sigma^{2}(x)\Big]K_{1}^{2}\bigg)+o(1)EK_1^2\Bigg\}\nonumber\\
&= \frac{EK_{1}^{2}}{nE^{2}K_{1}}(\sigma^{2}(x)+o(1))\simeq
\frac{\sigma^{2}(x)\xi_{2}}{n\varphi_{x}(\underline{h}\lambda)\xi_{1}^{2}%
}.\label{asyvar}%
\end{align}
Following similar arguments and using the latter assumption of B4, it can be readily shown that the covariance term is of smaller order than (\ref{asyvar}).

Noting (\ref{asybias}) and (\ref{asyvar}) we consider the normalized statistic $R^{*}_{nt}:=\sqrt{\varphi_x(\underline{h}\lambda)}\cdot R_{nt}$ and derive the limiting distribution of $(1/\sqrt{n})\cdot R_{nt}^{*}$. The usual procedure for establishing the asymptotic normality for the case of dependent observations is to employ Bernstein's blocking method; alternating blocks of two different sizes (hereafter referred to as the ``big'' and ``small'' blocks) are chosen to partition the index set $t=1,\ldots,n$. The argument proceeds by showing that the ``small blocks'' can be appropriately selected in such a way that they can be omitted in large sample, and then asymptotic independence can be obtained over the ``big blocks''. One may then apply the standard arguments involving the Lindeberg-Feller central limit theorem upon checking the Lindeberg conditions.

Below we shall only prove the autoregression case, where an additional step of mixing approximation is added to the standard regression case. The asymptotic normality for the regression case was established by Masry (2005) in a general context; we follow and briefly sketch his proof for the sake of completeness. Partition the index set $\{1,\ldots,n\}$ by $2k(=2k_n\rightarrow\infty)$ number of blocks of two different sizes that alternate (called the ``big-blocks'' and the ``small-blocks'', respectively), along with a single block (the ``last block'') that covers the remainder. The size of the alternating blocks is given by $a_n$ and $b_n$ respectively, where the one for the ``big-blocks'' $a_n$ is set to dominate that for the ``small-blocks'' $b_n$ in large sample, i.e. $b_n=o(a_n)$. More specifically, we take $k_n=\lfloor n/(a_n+b_n)\rfloor$ and $a_n=\lfloor \sqrt{n\varphi_x(\lambda\underline{h})}/q_n \rfloor$ where $q_n$ is a sequence of integer that goes off to infinity; it then clearly follows that $a_n/n\rightarrow0$ and $a_n/\sqrt{n\varphi_x(\lambda\underline{h})}\rightarrow0$. We also assume  $(n/a_n)\cdot\alpha^{*}(b_n)\rightarrow 0$, where $\alpha^{*}$ is the mixing coefficient for $R_{nt,(m)}^{*}$ below.

By construction above we can write $\sqrt{n}^{-1}\sum_{t=1}^{n}R_{nt}^{*}$ as the sum of the groups of big-blocks $\mathcal{B}$, small-blocks $\mathcal{S}$ and the remainder block $\mathcal{R}$ defined as
\begin{align*}
\mathcal{B}&:= \frac{1}{\sqrt{n}}\sum_{j=0}^{k-1}\Xi_{1,j}=\frac{1}{\sqrt{n}}\sum_{j=0}^{k-1}\left(\sum_{t=j(a+b)+1}^{j(a+b)+a}R_{nt}^{*}\right)\\
\mathcal{S}&:=\frac{1}{\sqrt{n}}\sum_{j=0}^{k-1}\Xi_{2,j}=\frac{1}{\sqrt{n}}\sum_{j=0}^{k-1}\left(\sum_{t=j(a+b)+a+1}^{(j+1)(a+b)}R_{nt}^{*}\right)\\
\mathcal{R}&:=\frac{1}{\sqrt{n}}\Xi_{3,j}=\frac{1}{\sqrt{n}}\left(\sum_{t=k(a+b)+1}^{n}R_{nt}^{*}\right).
\end{align*}

The aim is to show that the contributions from the small and the last remaining block are negligible, and that the big-blocks are asymptotically independent.

We first consider the big blocks $\mathcal{B}$, writing
\begin{align*}
\mathcal{B}=\sum_{j=0}^{k-1}\left(\sum_{t=j(a+b)+1}^{j(a+b)+a}R_{nt,(m)}^{*}\right)+\sum_{j=0}^{k-1}\left(\sum_{t=j(a+b)+1}^{j(a+b)+a}R_{nt,(m)}^{*}-\sum_{t=j(a+b)+1}^{j(a+b)+a}R_{nt}^{*}\right)=\mathcal{Q}_1+\mathcal{Q}_2,
\end{align*}
where $m$ is as in S2 and $R_{nt,(m)}^{*}=E(R_{nt}^{*}|Y_t,\ldots,Y_{t-m+1})$. As for the term $\mathcal{Q}_2$, write
\begin{align*}
\mathcal{Q}_2=&~\sum_{j=0}^{k-1}\sum_{t=j(a+b)+1}^{j(a+b)+a}\bigg(R_{nt,(m)}^{*}-R_{nt}^{*}\bigg)\\
=&~\sum_{j=0}^{k-1}\sum_{t=j(a+b)+1}^{j(a+b)+a}\frac{1}{EK_1}\bigg[K_{t}Y_{t}-K_{t,(m)}Y_{t}-\textrm{E}(K_{t}Y_{t})+\textrm{E}\big(K_{t,(m)}Y_{t}\big)\bigg]\\
&~~~~~~~-~\sum_{j=0}^{k-1}\sum_{t=j(a+b)+1}^{j(a+b)+a}\frac{1}{EK_1}\cdot m(x)\bigg[K_{t}
-K_{t,(m)}-\textrm{E}K_{t}
+\textrm{E}K_{t,(m)}\bigg]=\mathcal{I}_1+\mathcal{I}_2.
\end{align*}
Noting that the terms being summed up in $\mathcal{I}_1$ are the same as those in $R_2+R_3$ in p. 24 and that $R_{nt}^{*}$ is near epoch dependent on $Y_t$ 
it can be easily seen that $\sqrt{n}^{-1}\mathcal{Q}_2=o_p(1)$ by following similar arguments used before, and using the conditions on $a_n, b_n$ and $k_n$.
Further, the asymptotic independence of terms in $\mathcal{Q}_1$ holds by the Volkonskii-Rozanov inequality (see Fan and Yao (2003, p. 72)), and by the fact that $(n/a_n)\cdot\alpha^{*}(b_n)\rightarrow 0$.

Moving on to the small blocks, due to stationarity we have
\begin{align*}
&\textrm{Var}\left(\mathcal{S}\right)=\frac{1}{n}\textrm{Var}\left(\sum_{j=0}^{k-1}\sum_{t=j(a+b)+a+1}^{(j+1)(a+b)}R_{nt}^{*}\right)\\
&=\frac{1}{n}\sum_{j=0}^{k-1}\textrm{Var}\left(\sum_{t=j(a+b)+a+1}^{(j+1)(a+b)}R_{nt}^{*}\right)+\frac{1}{n}\mathop{\sum\sum}_{j\neq l}^{k-1}\textrm{Cov}\left(\sum_{t=j(a+b)+a+1}^{(j+1)(a+b)}R_{nt}^{*},\sum_{s=l(a+b)+a+1}^{(l+1)(a+b)}R_{ns}^{*}\right)\\
&=\frac{1}{n}\sum_{j=0}^{k-1}\bigg(b_n\textrm{Var}(R_{nt}^{*})+\sum_{t\neq l}^{b_n}\textrm{Cov}(R_{nt}^{*},R_{nl}^{*})\bigg)+\frac{1}{n}\mathop{\sum\sum}_{j\neq l}^{k-1}\sum_{i,j=1}^{b_n}\textrm{Cov}\big(R_{n,i+w_j}^{*},R_{n,r+w_l}^{*}\big)\\
&=Q_1+Q_2+Q_3.
\end{align*}
where $w_j=j(a+b)+a$. 

Regarding the first term, similar arguments used in deriving (\ref{asyvar}) yield
\begin{equation}
Q_1=\frac{1}{n}k_nb_n\frac{\big[\varphi_x(\underline{h}\lambda)^{1/2}\big]^2\sigma^2(x)\xi_2}{\varphi_x(\underline{h}\lambda)\xi_1^2}=\frac{k_nb_n\sigma^2(x)\xi_2}{n\xi_1^2}\longrightarrow0
\end{equation}
because $k_nb_n/n\sim b_n/(a_n+b_n)\rightarrow 0$. Now moving on to $Q_2$ and $Q_3$, the sum of covariances can be dealt with in the same manner as we did for the variance using 
(\ref{asyvar}), so $Q_2\rightarrow0$. 
Similarly for $Q_3$, implying $\textrm{Var}(\mathcal{S})\rightarrow 0$ as desired. Convergence result for the remainder $\mathcal{R}$ can be established similarly, and is bounded by $C(a_n+b_n)/n\rightarrow 0$.

Consequently, upon checking the Lindeberg conditions it follows that
\begin{equation}
\frac{\sum_{t} R_{nt}-ER_{nt}}{\sqrt{n\varphi_{x}(\lambda\underline{h}%
)^{-1}\sigma^{2}(x)\xi_{2}\xi_{1}^{-2}}} = \frac{{m}_{2}(x)-E\widehat{m}_{2}(x)-m(x)[\widehat{m}_{1}(x)-1]}{\Big(\sqrt{n\varphi_{x}(\lambda
\underline{h})}\Big)^{-1}\sqrt{\sigma^{2}(x)\xi_{2}\xi_{1}^{-2}}%
}\Longrightarrow N\big(0,1\big).\label{q39}%
\end{equation}

\noindent The rest of the proof concentrates on specification of the small ball probability:
\begin{align}
\varphi_{x}(\lambda\underline{h}) =&~P\big(\|H^{-1}(x-X)\|\leq
1\cdot\lambda\big)=P\big(X \in\mathcal{E}(x,\lambda\underline{h})\big)\notag\\
=&~P\bigg(\sum_{j=1}^{\infty
}j^{-2p}\big(x_{j}-X_{j}\big)^{2}\leq h^{2}\lambda^{2}\bigg)=P\Big(\|z-Z\|\leq h\lambda\Big)\nonumber\\
=&~\int_{B(0,h\lambda)}~dP_{z-Z}(u)=\int_{B(0,h\lambda)}p^{*}(u)~dP_Z(u)\simeq p^{*}(0)\cdot P(\|Z\|\leq h\lambda)\label{q41}
\end{align}
for any constant vector $z\in\ell_2$, where
$p^{*}(\cdot)$ is the Radon-Nikodym derivative of the induced probability measure $P_{z-Z}$ with respect to $P_Z$ (cf. Assumption B1); the explicit form of the derivative is not known in general. Nonetheless, in the special case of Gaussian process it is well-known that
\begin{equation}
P\Big(\|z-Z\|\leq \epsilon\Big)\simeq
P\Big(\|Z\|\leq \epsilon\Big)\exp\bigg\{-\frac{1}{2}\|\Gamma^{-1/2}z\|^{2}\bigg\}\quad\text{as}\quad\epsilon\rightarrow0,\label{zolo2}
\end{equation}
where the asymptotic equivalence is obtained by establishing the upper and lower bounds using Anderson's inequality (e.g. Proposition 2, Lewandowski et al. (1995)) and the Cameron-Martin theorem, respectively. See for instance Zolotarev (1988), or Li and Shao (2001) for detailed discussion.

Having reduced the problem of shifted small deviation probability to the non-shifted case, we can now specify the explicit rate of convergence upon substituting $r=h^{2}\lambda^{2}$, $A=2p$, and $a=2p/(2p-1)$ in
Proposition 4.1 of Dunker et al. (1998) for the i.i.d. case. In the dependent case, i.e. when the regressors satisfy Assumption D2, the small ball probability can be specified in view of Theorem 1 of Hong, Lifshits and Nazarov (2016), completing the proof.\hfill\mbox{\ \rule{.1in}{.1in}}
\bigskip

\subsection{Proof of Lemma 1 and 2}

Lemma 1 is a straightforward extension of Lemma 4.3 and 4.4 of Ferraty and
Vieu (2006), and hence is omitted. Lemma 2 can be shown by noting that for each $n$ the $\tau_{n}$%
-dimensional polyhedron $D:=\{w=(w_{i})_{i\leq\tau}\in\mathbb{R}^{\tau},
|w_{i}|\leq \lambda\}$ can be covered by $([2\lambda\sqrt{\tau}%
/\varepsilon+1])^{\tau }$ number of balls of radius $\varepsilon$, see Chat\'{e} and Courbage (1997), and then following the proof of Theorem 2 in Jia et al. (2003). \hfill%
\mbox{\
\rule{.1in}{.1in}} 

\subsection{Proof of Theorem 4}

As before, we start from the decomposition (\ref{decomp1}): 
\begin{align}
\widehat{m}(x)-m(x) = \frac{1}{\widehat{m}_{1}(x)}\Bigg(\Big[\widehat{m}%
_{2}(x)-E\widehat{m}_{2}(x)\Big]+\Big[E\widehat{m}_{2}(x)-m(x)\Big]-m(x)\Big[%
\widehat{m}_{1}(x)-1\Big]\Bigg).  \notag
\end{align}
Noting that the small deviation for $X=(X_{1},\ldots ,X_{\tau
},0,0,\ldots )$ is 
\begin{equation*}
\varphi _{x}(\lambda \underline{h})=P\Bigg(\sum_{j=1}^{\tau
}j^{-2p}X_{j}^{2}\leq h^{2}\Bigg)\simeq P\Bigg(\sum_{j=1}^{\infty
}j^{-2p}X_{j}^{2}\leq h^{2}+\tau ^{-2p+1}\Bigg),
\end{equation*}
the proof reduces to showing three steps, where in the first step we show 
\begin{equation}
\sup_{x\in\mathcal{S}_{\tau}}\Big|\widehat{m}_{2}(x)-E\widehat{m}_{2}(x)\Big|%
= O_{a.s.}\Bigg(\sqrt{\frac{(\log n)^{2}}{n\varphi_{x}(\lambda\underline{h})}%
}~\Bigg).  \label{q43}
\end{equation}
We cover the set $\mathcal{S}_{\tau}$ defined in (\ref{q18}) by $L=L(%
\mathcal{S}_{\tau},\eta)$ number of balls of radius $\eta$ denoted $I_{k}$,
each of which is centred at $x_{k}$, for $k=1,...,L$. i.e. $S_{\tau}
\subset\bigcup _{k=1}^{L_{n}} B(x_{k,n},\eta)$. Then it follows that 
\begin{align}
\sup_{x\in\mathcal{S}_{\tau}}&\Big|\widehat{m}_{2}(x)-E\widehat{m}_{2}(x)%
\Big| = \max_{1\leq k\leq L_{n}}\sup_{x\in I_{k}\cap\mathcal{S}_{\tau}}\Big|%
\widehat {m}_{2}(x)-E\widehat{m}_{2}(x)\Big|  \notag \\
& =\max_{1\leq k\leq L_{n}}\sup_{x\in I_{k}\cap\mathcal{S}_{\tau}}\Big|%
\widehat {m}_{2}(x)-\widehat{m}_{2}(x_{k})+\widehat{m}_{2}(x_{k})-E\widehat{m%
}_{2}(x_{k})+E\widehat {m}_{2}(x_{k})-E\widehat{m}_{2}(x)\Big|  \notag \\
& \leq\max_{1\leq k\leq L_{n}}\sup_{x\in I_{k}\cap\mathcal{S}_{\tau}}\Big|%
\widehat{m}_{2}(x)-\widehat{m}_{2}(x_{k})\Big|+\max_{1\leq k\leq
L_{n}}\sup_{x\in I_{k}\cap\mathcal{S}_{\tau}}\Big|E\widehat{m}_{2}(x_{k})-E%
\widehat{m}_{2}(x)\Big|  \notag \\
& ~~~+\max_{1\leq k\leq L_{n}}\Big|\widehat{m}_{2}(x_{k})-E\widehat{m}%
_{2}(x_{k})\Big| ~=: R_{1}+R_{2}+R_{3}.  \label{q44}
\end{align}

\noindent We first consider $R_{1}$: 
\begin{align}
R_{1} & =\max_{1\leq k\leq L_{n}}\sup_{x\in I_{k}\cap S_{\tau}}\Big|%
\widehat {m}_{2}(x)-\widehat{m}_{2}(x_{k})\Big|  \notag \\
& =\max_{1\leq k\leq L_{n}}\sup_{x\in I_{k}\cap S_{\tau}}\Bigg|\frac {1}{%
nEK_{1}}\sum_{t=1}^{n}Y_{t}K\Big(\|H^{-1}(x-X_{t})\|\Big)-Y_{t}K\Big(%
\|H^{-1}(x_{k}-X_{t})\|\Big)\Bigg|  \notag \\
& \leq\max_{1\leq k\leq L_{n}}\sup_{x\in I_{k}\cap S_{\tau}}\frac{C}{%
n\varphi_{x}(\lambda\underline{h})}\sum_{t=1}^{n}\big|Y_{t}K_{t}-Y_{t}K_{t,k}%
\big|\cdot\mathbf{1}_{\mathcal{E}(x,\lambda\underline{h})}(X_t)\cdot\mathbf{1%
}_{\mathcal{E}(x_{k},\lambda\underline{h})}(X_t),  \notag
\end{align}
where $K_{t,k}:=K(\|H^{-1}(x_k-X_t)\|)$. Now, because the kernel function is
assumed to be Lipschitz continuous by Assumption C2, on choosing $\eta=\log
n/n$ we get 
\begin{align}
R_{1} \leq\frac{1}{n}\sum_{t=1}^{n}\frac{|Y_t|}{\varphi_{x}(\underline{h}%
\lambda)}\eta h^{-1}\mathbf{1}_{\mathcal{E}(x,\lambda\underline{h})\cup%
\mathcal{E}(x_{k},\lambda\underline{h})}(X_{t}) =:\frac{1}{n}%
\sum_{t=1}^{n}J_{t}, \notag
\end{align}
where $J_t$ is $\alpha$-mixing under both assumptions S1' and S2'. Then for some $\delta>0$ 
\begin{equation}
E|J_t|^{2+\delta}\leq\frac{E|Y_t|^{2+\delta}\eta^{2+\delta}}{%
\varphi_{x}(\lambda\underline{h})^{2+\delta}h^{2+\delta}}\varphi_{x}(\lambda%
\underline{h})\leq C\left(\frac{\log n}{n\varphi_x(\lambda\underline{h})h}%
\right)^{1+\delta}\times\left(\frac{\log n}{nh}\right)\leq C^{\prime }
\label{q46}
\end{equation}
by Assumption A4 and C1, which suggests that for any positive $t$ 
\begin{equation}
P(|J_t|>t)\leq\frac{E|J_t|^{2+\delta}}{t^{2+\delta}}\leq C^{\prime}t^{
-(2+\delta)}.  \label{FukNagaev}
\end{equation}
Also, by Assumption A6 we have 
\begin{equation}
E|J_t|\leq\frac{E(E(|Y_t||X))\eta}{\varphi_{x}(\lambda\underline{h})h}%
\varphi_{x}(\lambda\underline{h})\leq\frac{C\eta}{h}.  \label{bbias}
\end{equation}
By Lemma 2 we can specify the Kolmogorov's entropy for $S_{\tau}$ with $%
\eta=\log n/n$: 
\begin{equation*}
\log L\bigg(S,\frac{\log n}{n}\bigg)=C\log\Bigg[\Bigg(\frac{2\lambda n}{\sqrt{\log n}}+1\Bigg)^{\log n}~\Bigg]\sim\log
n\times\log\Bigg[\frac{2\lambda n}{\sqrt{\log n}}\Bigg]
\end{equation*}
for sufficiently large $n$ and $\lambda$, implying that the order of
Kolmogorov's $\frac{\log n}{n}$ entropy is 
\begin{equation}
O\bigg(\log L\bigg(S_{\tau},\frac{\log n}{n}\bigg)\bigg)=O\bigg((\log
n)^{2}-\log n[\log\log n]\bigg)=O\bigg(\big(\log n\big)^{2}\bigg).
\label{q47}
\end{equation}
Now, with (\ref{FukNagaev}) we may apply the Fuk-Nagaev 
inequality\footnote{%
See Rio (2000, p. 87) for details; the exact bounds of the constants are
specified.} with $r=(\log n)^{2}$ and $\varepsilon=\varepsilon_{0}[\log L%
\big(S,\frac{\log n}{n}\big)/(n\varphi(\lambda\underline{h}))]^{1/2}$  for some positive constant $\varepsilon_0$. Then since
\begin{equation*}
s_n^2:=\sum_{t=1}^{n}\sum_{s=1}^{n}\text{Cov}\left(J_t,J_s\right)\leq C\left(%
\frac{1}{\varphi(\lambda\underline{h})}\frac{(\log n)^2}{n^2h^2}-\frac{(\log
n)^2}{n^2h^2}\right)=O\left(\frac{n\log n}{\varphi(\lambda\underline{h})}%
\right),
\end{equation*}
and because an exponentially $\alpha$-mixing sequence is arithmetrically $\alpha$-mixing with some mixing rate $k>0$, it follows that 
\begin{align}
P\Bigg(\frac{1}{n}\bigg|&\sum_{t=1}^{n}J_{t}-EJ_{1}\bigg| >\varepsilon _{0}%
\sqrt{\frac{\log L\big(S,\frac{\log n}{n}\big)}{n\varphi(\lambda \underline{h%
})}} \Bigg)  \notag \\
& \leq~ 4\bigg(1+\frac{n^{2}\varepsilon_{0}^{2}\log L\big(S,\frac{\log n}{n}%
\big)}{16(\log n)^{2}s^{2}_{n}n\varphi(\lambda\underline{h})}\bigg)^{-\frac{%
(\log n)^{2}}{2}}+\frac{C^{\prime }\cdot n}{(\log n)^2}\Bigg(\frac {(\log
n)^4 n\varphi(\lambda\underline{h})}{n^2\varepsilon_0^2\log L\big(S,\frac{%
\log n}{n}\big)}\Bigg)^{\frac{(k+1)(2+\delta)}{2(2+\delta+k)}}  \notag \\
& =~ 4\bigg(1+\frac{n}{16s_n^2\varphi(\lambda\underline{h})}%
\varepsilon_{0}^{2}\bigg)^{-\frac{(\log n)^{2}}{2}}+\frac{C^{\prime }\cdot n%
}{(\log n)^2}\left( \frac{\varphi(\lambda\underline{h})}{n\varepsilon_0^2}
\right)^{\frac{2k+k\delta+2+\delta}{4+2\delta+2k}}(\log n)^{\frac{%
(k+1)(2+\delta)}{2+\delta+k}}  \notag \\
& \leq~ 4\exp\bigg(-\frac{\varepsilon_{0}^{2}(\log n)^2n}{%
32s_n^2\varphi(\lambda\underline{h})}\bigg)+ C^{\prime} n^{-\frac{%
k\delta-(\delta+2)}{2(2+\delta+k)}}(\log n)^{\frac{k\delta-(\delta+2)}{%
2+\delta+k}}\varphi(\lambda\underline{h})^{\frac{2k+k\delta+2+\delta}{%
2(2+\delta+k)}}  \notag \\
& \leq~ 4\exp\bigg(-\frac{\varepsilon_{0}^{2}\log n}{32}\bigg)+ C^{\prime
}\left(\frac{\log n}{n^2}\right)^{\frac{k\delta-(\delta+2)}{2+\delta+k}}
\label{convp}
\end{align}
by choosing $k$ sufficiently large so that $k\delta>\delta+2$ and Taylor
expansion of $\log(1+\epsilon)$ for sufficiently small $\epsilon>0$. 
Now summing up over $n=1$ upto infinity, we see that
\begin{align}
\sum_{n=1}^{\infty}P\Bigg(&\frac{1}{n}\bigg|\sum_{i=1}^{n}J_{i}-EJ_{1}\bigg|%
>\varepsilon_{0}\sqrt{\frac{(\log n)^2}{n\varphi(\lambda\underline{h})}} %
\Bigg)\notag\\
&\leq C\sum_{n=1}^{\infty}\exp\bigg(-\frac{\varepsilon_{0}^{2}\log n%
}{32}\bigg)+\sum_{n=1}^{\infty} \left(\frac{\log n}{n^2}\right)^{\frac{%
k\delta-(\delta+2)}{2+\delta+k}}<\infty  \notag
\end{align}
by choosing $\varepsilon_0$ sufficiently big (i.e. any finite real number
strictly bigger than $4\sqrt{2}$). 

Hence by (\ref{bbias}) and by the Borel-Cantelli lemma we have 
\begin{align}
R_{1}&=\max_{1\leq k\leq L_{n}} \sup_{x\in I_{k}\cap \mathcal{S}_{\tau }}%
\Big|\widehat{m}_{2}(x)-\widehat{m}_{2}(x_{k})\Big|\leq O\bigg(\frac{\eta }{h%
}\bigg)+O_{a.s.}\Bigg(\sqrt{\frac{(\log n)^{2}}{n\varphi _{x}(\lambda 
\underline{h})}}\Bigg)  \notag \\
&=O\Bigg(\sqrt{\frac{(\log n)^{2}}{n^{2}h^{2}}}\Bigg)+O_{a.s.}\Bigg(\sqrt{%
\frac{(\log n)^{2}}{n\varphi _{x}(\lambda \underline{h})}}\Bigg)=O_{a.s}%
\Bigg(\sqrt{\frac{(\log n)^{2}}{n\varphi _{x}(\lambda \underline{h})}}~\Bigg)%
.  \label{q49}
\end{align}%
\noindent For the second term $R_{2}$, following similar arguments
as above we can show that 
\begin{equation}
R_{2}\leq \max_{1\leq k\leq L_{n}}\sup_{x\in I_{k}\cap \mathcal{S}_{\tau }}E%
\big|\widehat{m}_{2}(x)-\widehat{m}_{2}(x_{k})\big|=O\Bigg(\sqrt{\frac{(\log
n)^{2}}{n\varphi _{x}(\lambda \underline{h})}}~\Bigg).\label{q50}
\end{equation}%
Next we study the last component: 
\begin{equation}
R_{3}=\max_{1\leq k\leq L_{n}}\big|\widehat{m}_{2}(x_{k})-E\widehat{m}%
_{2}(x_{k})\big|=:\max_{1\leq k\leq L_{n}}\big|W_{n}(x_{k})\big|  \label{q51}
\end{equation}%
where 
\begin{align}
W_{n}(x)& =\widehat{m}_{2}(x)-E\widehat{m}_{2}(x)=\frac{1}{nEK_{1}}%
\sum_{t=1}^{n}\Big[Y_{t}K_{t}-EY_{t}K_{t}\Big]  \notag \\
& =\frac{1}{n}\sum_{t=1}^{n}\bigg[\frac{Y_{t}K_{t}}{EK_{1}}-\frac{EY_{t}K_{t}%
}{EK_{1}}\bigg]=:\frac{1}{n}\sum_{t=1}^{n}Q_{t}  \notag
\end{align}%
as defined in (\ref{mmp}), and by elementary arguments 
\begin{align}
P\bigg( \max_{1\leq k\leq L_{n}}\vert \widehat{m}_{2}(x_{k})&-E\widehat{m%
}_{2}(x_{k})\vert >\varepsilon \bigg)\leq L_{n}\cdot \sup_{x\in 
\mathcal{S}}P\big(\left\vert W_{n}(x)\right\vert >\varepsilon \big).\label{Ln}
\end{align}
As we did before in the pointwise case, we must separately consider the
cases of regression and autoregression because the asymptotic arguments to
follow depends upon the temporal dependence structure of $Q_{t}$.

In the regression case, we use the exponential inequality of Bosq (1996, Theorem 1.3.2) for $\alpha$-mixing sequence. We note that $|Q_{t}|\leq
C/\varphi_{x}(\underline{h}\lambda)=:b$ for all $t$, and that $\sigma^{2}(m):=p\cdot
\textrm{Var}(Q_{t})=O(p/\varphi(\underline{h}\lambda))$ (where $p=n/(2q)$ and $q=\log n\sqrt{n}/\sqrt{\varphi}$) by the Cauchy-Schwarz inequality and assumption A4. Hence, it follows that
\[
v^{2}(m)=\frac{2}{p^{2}}\sigma^{2}(m)+\frac{b\varepsilon}{2}\leq\frac
{Cq}{n\varphi_{x}(\underline{h}\lambda)}+\frac{C\varepsilon}{\varphi
_{x}(\underline{h}\lambda)}\leq\frac{C^{\prime}\varepsilon}{\varphi
_{x}(\underline{h}\lambda)},
\]
where $\varepsilon=\varepsilon_0\sqrt{\log L_n/(n\varphi)}$, and by assumption S1 that
\begin{align}
P\Bigg(\Big|\widehat{m}_{2}(x)&-E\widehat{m}_{2}
(x)\Big|>\varepsilon_0\sqrt{\frac{\log L\left(S,\frac{\log n}{n}\right)}{n\varphi(\lambda\underline{h})}}~\Bigg)\notag\\
&\leq4\exp\bigg\{-\frac{\varepsilon^{2}q}{8v^{2}(m)}\bigg\}+22\bigg(1+\frac{4b}{\varepsilon}\bigg)^{1/2}q\alpha\left(
\left[ \frac{n}{2q}\right] \right) \nonumber\\
&\leq 4\exp\left\{-\frac{\varepsilon_0 q\varphi\sqrt{\log L_n} }{8\sqrt{n\varphi}}\right\}+22\bigg(1+\frac{4\sqrt{n\varphi}}{\varphi \log n}\bigg)^{1/2}\frac{\log n\sqrt{n}}{\sqrt{\varphi}}\alpha\left(
\left[ \frac{\sqrt{n\varphi}}{2\log n}\right] \right) \nonumber\\
&\leq 4\exp\left\{-\frac{\varepsilon_0 \log L_n }{8}\right\} \leq CL_n^{-\varepsilon_0/8},
\end{align}
due to exponential decay of $\alpha(\cdot)$; with (\ref{Ln}) the desired result follows by choosing $\varepsilon_0$ large enough. 

In the autoregression case (i.e. under assumption D2), using the mixing approximation
techniques adopted in Section 4.1. and following the arguments in (\ref{autor}), (\ref{autor1}) and (\ref{p10}), it can be readily shown that the same conclusion can be derived; that is:
\begin{equation}
R_{3}=\max_{1\leq k\leq L_{n}}\bigg|\widehat{m}_{2}(x_{k})-E\widehat{m}%
_{2}(x_{k})\bigg|=O_{a.s.}\Bigg(\sqrt{\frac{(\log n)^{2}}{n\varphi
_{x}(\lambda \underline{h})}}~\Bigg).  \label{q54}
\end{equation}%
Now returning back to where we started, viewing $\widehat{m}_{1}(x)$ as a
special case of $\widehat{m}_{2}(x)$ with $Y_{t}=1$ $\forall t$, we can
repeat the above procedure, yielding (since $E\widehat{m}_{1}(x)=1$)
\begin{equation}
\sup_{x\in \mathcal{S}_{\tau }}\Big|\widehat{m}_{1}(x)-1\Big|=O_{a.s.}\Bigg(%
\sqrt{\frac{(\log n)^{2}}{n\varphi _{x}(\lambda \underline{h})}}~\Bigg).
\label{q55}
\end{equation}%
Finally the proof is complete in view of (\ref{q43}), (\ref{q49}), (\ref{q50}), (\ref{q54}), (\ref{q55}), and the contribution from the bias component.
\hfill \mbox{\ \rule{.1in}{.1in}} 


\bigskip

\begin{thebibliography}{99}
\bibitem{} de Acosta, A., 1983. \emph{Small deviations in the functional
central limit theorem with applications to functional laws of the iterated
logarithm}. Annals of Probability, 11(1), 78-101.

\bibitem{} Andrews, D. W., 1995. \emph{Nonparametric kernel estimation for
semiparametric models}. Econometric Theory, 11(3), 560-586.

\bibitem{} Bierens, H. J., 1983. \emph{Uniform consistency of kernel
estimators of a regression function under generalized conditions}. Journal
of the American Statistical Association, 78(383), 699-707.

\bibitem{} Bierens, H. J., 1987. \emph{Kernel estimators of regression
functions}. In Advances in econometrics: Fifth world congress, 1, 99-144.

\bibitem{} Bingham, N. H., Goldie, C. M., Teugels, J. L., 1987. \emph{%
Regular Variation}. Cambridge University Press.

\bibitem{} Bochner, S., 1955. \emph{Harmonic Analysis and the Theory of
Probability}. University of California Press, Berkeley and Los Angeles.

\bibitem{} Bosq, B., 1996. \emph{Nonparametric statistics for stochastic
processes: estimation and prediction}. Springer-Verlag, New York.

\bibitem{} Bradley, R. C., 2005. \emph{Basic properties of strong mixing
conditions. A survey and some open questions}. Probability Surveys, 2(2),
107-144.

\bibitem{} Bulinskii, A. V., Shiryaev, A. N., 2003. \emph{Theory of Random
Processes} (in Russian). Fizmatlit.


\bibitem {} Chat\'{e}, H., Courbage, M. (ed.), 1997. \emph{Special issue on lattice dynamics}. Physica D, 103, 1-612.

\bibitem{} Davidson, J., 1994. \emph{Stochastic Limit Theory: An
Introduction for Econometricians: An Introduction for Econometricians.}
Oxford University Press.

\bibitem {} Davydov, Y. A., 1968. \emph{Convergence of distributions generated by stationary stochastic processes}. Theory of Probability and Applications, 13(4), 691-696.

\bibitem{} Delsol, L., 2009. \emph{Advances on asymptotic normality in
non-parametric functional time series analysis}. Statistics, 43(1), 13-33.

\bibitem{} Devroye, L. P., 1978. \emph{The uniform convergence of the
Nadaraya-Watson regression function estimate}. Canadian Journal of
Statistics, 6(2), 179-191.

\bibitem{} Devroye, L., 1981. \emph{On the almost everywhere convergence of
nonparametric regression function estimates}. Annals of Statistics, 9(6),
1310-1319.

\bibitem{} Doukhan, P., 1994. \emph{Mixing}. Springer, New York.

\bibitem{} Doukhan, P., Wintenberger, O., 2008. \emph{Weakly dependent
chains with infinite memory}. Stochastic Processes and their Applications,
118(11), 1997-2013.

\bibitem{} Duflo, M., 1997. \emph{Random iterative models}. Springer-Verlag,
Berlin.

\bibitem{} Dunker, T., Lifshits, M. A., Linde, W., 1998. \emph{Small
deviation probabilities of sums of independent random variables}. In:
Eberlein, E. (ed.), High dimensional probability, volume 43 of Progress in
Probability., Birkhauser, Basel, 59-74.

\bibitem{} Fan, J., Masry, E., 1992. \emph{Multivariate regression
estimation with errors-in-variables: asymptotic normality for mixing
processes}. Journal of Multivariate Analysis, 43(2), 237-271.

\bibitem{} Fan, J., 1990. \emph{A remedy to regression estimators and
nonparametric minimax efficiency}. Department of Statistics, University of
North Carolina at Chapel Hill.

\bibitem{} Fan, J., Yao, Q., 2005. \emph{Nonlinear Time Series:
Nonparametric and Parametric Methods}. Springer, New York.

\bibitem{} Feller, W., 1971. \emph{Introduction to Probability Theory and
Its Applications}, Vol. 2. Wiley, New York.

\bibitem{} Ferraty, F., Laksaci, A., Tadj, A., Vieu, P., 2010. \emph{Rate of
uniform consistency for nonparametric estimates with functional variables}.
Journal of Statistical Planning and Inference, 140, 335-352.

\bibitem{} Ferraty, F., Romain, Y., 2010. \emph{The Oxford Handbook of
Functional Data Analysis}. Oxford University Press.

\bibitem{} Ferraty, F., Vieu, P., 2002. \emph{The functional nonparametric
model and application to spectrometric data}. Computational Statistics, 17,
545-564.

\bibitem{} Ferraty, F., Vieu, P., 2006. \emph{Nonparametric functional data
analysis: Theory and Practice}. Springer, New York.

\bibitem{} Fuk, D. K., Nagaev, S. V., 1971. \emph{Probability inequalities
for sums of independent random variables}. Theory of Probability and its
applications, 16, 643-660.

\bibitem{} G\"{o}tze, F., Hipp, C., 1994. \emph{Asymptotic distribution of
statistics in time series}. Annals of Statistics, 22(4), 2062-2088.

\bibitem{} Greblicki, W., Krzyzak, A., 1980. \emph{Asymptotic properties of
kernel estimates of a regression function}. Journal of Statistical Planning
and Inference, 4(1), 81-90.

\bibitem{} Geenens, G., 2011. \emph{Curse of dimensionality and related issues in nonparametric functional regression}. Statistics Surveys, 5, 30-43.


\bibitem{} H\"{a}dle, W. K., 1990. \emph{Applied Nonparametric Regression}.
(Vol. 27). Cambridge University Press.




\bibitem{} Hong, S. Y., Lifshits, M., Nazarov. A., 2015. \emph{Small
Deviations for Dependent Sequences}. Preprint available at: \texttt{%
http://arxiv.org/abs/1511.05370}.

\bibitem{} Ibragimov, I.A., Linnik., Y. B., 1971. \emph{Independent and
Stationary Sequences of Random Variables}. Wolters-Noordhoff, Groningen.

\bibitem{} Jia, Q., Zhou, S., Yin. F., 2003. \emph{Kolmogorov entropy of
global attractor for dissipative lattice dynamical systems}. Journal of
Mathematical Physics, 44, 5804-5801.

\bibitem{} Karamata, J., 1933. \emph{Sur un mode de croissance regulire.
Theoremes fondamentaux} (in French). Bulletin de la Societe Mathematique de
France, 61, 55-62.


\bibitem{} Lewandowski, M., Ryznar, M., 1995. \emph{Anderson inequality is
strict for Gaussian and stable measures}. Proceedings of the American
Mathematical Society, 123(12), 3875-3880.

\bibitem{} Li, W. V., Shao, Q. M., 2001. \emph{Gaussian processes:
inequalities, small ball probabilities and applications}. Handbook of
Statistics, 19, 533-597.

\bibitem{} Lifshits, M. A., 1997. On the lower tail probabilities of some
random series. Annals of Probability, 25(1), 424-442.

\bibitem{} Linton, O. B., Sancetta, A., 2009. \emph{Consistent estimation of
a general nonparametric regression function in time series}. Journal of
Econometrics, 152, 70-78.

\bibitem{} Lu, Z., 2001. \emph{Asymptotic normality of kernel density
estimators under dependence}. Annals of the Institute of Statistical
Mathematics, 53(3), 447-468.


\bibitem{} Mas, A., 2012. \emph{Lower bound in regression for functional
data by small ball probability representation in Hilbert space}, Electronic
Journal of Statistics, 6, 1745-1778.

\bibitem{} Masry, E., 2005. \emph{Nonparametric regression estimation for
dependent functional data: asymptotic normality}. Stochastic Processes and
their Applications, 115, 155-177.


\bibitem{} Nadaraya, E. A., 1964. \emph{On estimating regression}. Theory of
Probability \& Its Applications, 9(1), 141-142.

\bibitem{} Nadaraya, E. A., 1970. \emph{Remarks on non-parametric estimates
for density functions and regression curves}. Theory of Probability \& Its
Applications, 15(1), 134-137.

\bibitem{} Pagan, A., Ullah, A., 1988. \emph{The econometric analysis of
models with risk terms}. Journal of Applied Econometrics, 3(2), 87-105.

\bibitem{} Parzen, E., 1962. \emph{On estimation of a probability density
function and mode}. Annals of Mathematical Statistics, 33(3), 1065-1076.

\bibitem{} Phillips, P. C., Park, J. Y., 1998. \emph{Nonstationary density
estimation and kernel autoregression}. Cowles Foundation for Research in
Economics.

\bibitem{} Ramsay, J. O., Silverman, B. W., 2002. \emph{Applied functional
data analysis: methods and case studies}. Springer, New York.

\bibitem{} Rio, E., 2000. \emph{Theorie asymptotique des processus al\'{e}
atoires faiblement dependants} (French) Springer, Maththematiques and
Applications, 31.

\bibitem{} Robinson, P. M., 1983. \emph{Nonparametric estimators for time
series.} Journal of Time Series Analysis, 4(3), 185-207.

\bibitem{} Rosenblatt, M., 1956. \emph{A central limit theorem and a strong
mixing condition}. Proceedings of the National Academy of Sciences, 42(1),
43-47.

\bibitem{} Roussas, G. G., 1989. \emph{Consistent regression estimation with
fixed design points under dependence conditions}. Statistics \& Probability
Letters, 8(1), 41-50.

\bibitem{} Roussas, G. G., 1990. \emph{Nonparametric regression estimation
under mixing conditions}. Stochastic Processes and Their Applications,
36(1), 107-116. 


\bibitem{} Schuster, E. F., 1972. \emph{Joint asymptotic distribution of the
estimated regression function at a finite number of distinct points}. Annals
of Mathematical Statistics, 43(1), 84-88.

\bibitem{} Skorohod, A. B., 1967. \emph{On the densities of probability
measures in functional spaces}. Proc. Fifth Berkeley Sympos. Math. Statist.
and Probability (Berkeley, Calif., 1965/66) Vol. II: Contributions to
Probability Theory, 1, 163-182. University of California Press, Berkely, CA.

\bibitem{} Stone, C., 1980. \emph{Optimal rates of convergence for
nonparametric estimators}. Annals of Statistics, 8(6), 1348-1360.

\bibitem{} Stone, C., 1982. \emph{Optimal global rates of convergence for
nonparametric regression}. Annals of Statistics, 10(4), 1040-1053.



\bibitem{} Sytaya, G. N., 1974. \emph{Certain asymptotic representations for
a Gaussian measure in Hilbert space}. Theory of Random Process, 2, 93-104 (in Russian).



\bibitem{} Watson, G. S., 1964. \emph{Smooth regression analysis}, Sankhya
Ser. A, 26, 359-372.

\bibitem{} Wu, W. B., 2011. \emph{Asymptotic theory for stationary processes}%
. Statistics and Its Interface, 0, 1-20.

\bibitem{} Zolotarev, V. M., 1988. \emph{Asymptotic behavior of the Gaussian
measure in $\ell_{2}$}. Journal of Soviet Mathematics, 24, 2330-2334.
\end{thebibliography}
\end{document}